\documentclass[a4paper,11pt]{article}
\pdfoutput=1 

\usepackage{jcappub} 
\usepackage{aasmacros}
\newcommand*\ttvar[1]{\texttt{\expandafter\dottvar\detokenize{#1}\relax}}
\newcommand*\dottvar[1]{\ifx\relax#1\else
  \expandafter\ifx\string_#1\string_\allowbreak\else#1\fi
  \expandafter\dottvar\fi}

\usepackage[T1]{fontenc} 
\usepackage{physics}
\usepackage{multirow}
\usepackage[export]{adjustbox}

\newcommand\ChangeRT[1]{\noalign{\hrule height #1}}

\usepackage[scientific-notation=true]{siunitx}
\usepackage{lipsum}

\usepackage[parfill]{parskip}

\newcommand\deepsphere{\textsc{DeepSphere}}
\newcommand\emuname{\textsc{KiDS-cGAN}}

\defcitealias{original_cGAN}{P20}

\usepackage{caption}
\usepackage{subcaption}
\usepackage{enumerate}

\title{\boldmath A tomographic spherical mass map emulator of the KiDS-1000 survey using conditional generative adversarial networks}

\author[a,1]{Timothy Wing Hei Yiu,\note{Corresponding author.}}
\author[a]{Janis Fluri,}
\author[a,b]{Tomasz Kacprzak}

\affiliation[a]{Institute for Particle Physics and Astrophysics, ETH Zürich,\\Wolfgang-Pauli-Strasse 27, CH-8093 Zürich, Switzerland}
\affiliation[b]{Swiss Data Science Center, Paul Scherrer Institute\\Forschungsstrasse 111, 5232 Villigen PSI, Switzerland}

\emailAdd{yiu@astron.nl}
\emailAdd{jafluri@phys.ethz.ch}
\emailAdd{tomaszk@phys.ethz.ch}

\compress
\abstract{
Large sets of matter density simulations are becoming increasingly important in large-scale structure cosmology.
Matter power spectra emulators, such as the Euclid Emulator and CosmicEmu, are trained on simulations to correct the non-linear part of the power spectrum.
Map-based analyses retrieve additional non-Gaussian information from the density field, whether through human-designed statistics such as peak counts, or via machine learning methods such as convolutional neural networks.
The simulations required for these methods are very resource-intensive, both in terms of computing time and storage.
This creates a computational bottleneck for future cosmological analyses, as well as an entry barrier for testing new, innovative ideas in the area of cosmological information retrieval.
Map-level density field emulators, based on deep generative models, have recently been proposed to address these challenges.
In this work, we present a novel mass map emulator of the KiDS-1000 survey footprint, which generates noise-free spherical maps in a fraction of a second.
It takes a set of cosmological parameters $(\Omega_M, \sigma_8)$ as input and produces a consistent set of 5 maps, corresponding to the KiDS-1000 tomographic redshift bins.
To construct the emulator, we use a conditional generative adversarial network architecture and the spherical convolutional neural network \deepsphere, and train it on N-body-simulated mass maps.
We compare its performance using an array of quantitative comparison metrics: angular power spectra $C_\ell$, pixel/peaks distributions, $C_\ell$ correlation matrices, and Structural Similarity Index.
Overall, the average agreement on these summary statistics is $<10\%$ for the cosmologies at the centre of the simulation grid, and degrades slightly on grid edges.
However, the quality of the generated maps is worse at high negative $\kappa$ values or large scale, which can significantly affect summaries sensitive to such observables.
Finally, we perform a mock cosmological parameter estimation using the emulator and the original simulation set.
We find good agreement in these constraints, for both likelihood and likelihood-free approaches.
The emulator is available at \href{https://tfhub.dev/cosmo-group-ethz/models/kids-cgan/1}{tfhub.dev/cosmo-group-ethz/models/kids-cgan}.
}

\begin{document}
\maketitle
\flushbottom


\section{Introduction} \label{sec:intro}

The structure of the matter density field is an important source of information for testing cosmological models. 
Large-scale weak lensing surveys, such as the Kilo-Degree Survey (KiDS), the Dark Energy Survey (DES), and the Hyper Suprime-Cam (HSC), have made multiple measurements of the matter density $\Omega_M$, clustering strength $\sigma_8$, dark energy equation of state $w$ \cite{kids1000_shear,Amon2021desy3,Secco2021desy3,Hamana2020hsc}, the Hubble constant $H_0$ \cite{DES2018h0}, and tested extension to the standard cosmological $\Lambda$CDM model \cite{Troester2021kids1000,DES20219extended}. 
Stage-IV large-scale structure surveys such as those carried out by the Vera C. Rubin Observatory and Euclid will soon begin observations, and will greatly increase the precision of the measurement of these parameters.

While the density field is close to Gaussian at large scales, the non-linear growth of structure dominates the intermediate and small scales.
Accurate predictions of these scales require dark matter N-body simulations, whether to tune parameters of semi-analytic models (\textsc{HaloFit} \cite{Takahashi2012halofit}, \textsc{HMcode} \cite{Mead2021hmcode}) or to construct the emulators directly (Euclid Emulator \cite{euclid2019euclid}, \textsc{Bacco} \cite{Angulo2020Bacco}, \textsc{CosmicEmu} \cite{Lawrence2017cosmicemu}). The reliance on emulator approaches is expected to rise with increasing requirements on prediction precision \cite{Martinelli2021nonlinear}.
These emulators are typically focused on predicting summary statistics of the density field, most commonly the 3D matter power spectrum $P_k$.
The predicted $P_k$ signal is then used to calculate the 2-point functions (the power spectrum $C_\ell$ or correlation function $\xi_{\pm}$, and others) of different large-scale structure (LSS) observables, such as cosmic shear and galaxy clustering.
The $P_k$ distribution is then modelled as a multivariate Gaussian, with covariance calculated at a fixed cosmology \cite{Eifler2009covariance}.

The 2-point functions capture only the Gaussian part of the non-linear density field, and thus ignore information beyond Gaussian. The non-Gaussian information contains a significant cosmological signal, and it responds differently to astrophysical and measurement systematics \cite{zurcher2021,dominik_DES,Marian2013peaks,HarnoisDeraps2021peaks,fluri2018cosmological}, especially in the intrinsic galaxy alignments and redshift calibration sector \cite{dominik_DES}.
Extracting this information has gained significant interest recently, with multiple techniques proposed for this task.
Pre-defined statistics, such as peak counts \cite{Dietrich2010peaks}, Minkowski functionals \cite{Petri2013minkowski}, three-point functions \cite{Takada2003three}, and map moments \cite{Gatti2020moments}, have been used to make cosmological measurements \citep{Fu2014three,zurcher2021,Gatti2021moments}.
Furthermore, machine learning approaches that automatically extract informative features, such as deep convolutional neural networks \cite{fluri2018cosmological,gupta2018non,Kaushal2021necola}, have recently been used to make first parameter measurements \cite{kids450_dl} and to extract information at map level (see \cite{ML_Peel2019,ML_Jeffrey2021,ML_Makinen2021,ML_Villaescusa2020} for examples).

These methods require a fully numerical LSS theory prediction, capturing both the map signal and its variability.
This prediction is typically done by creating simulations on fixed grids of cosmological parameters.
Recent measurements with DES data used a large simulation set called the \textsc{DarkGridV1}, which consisted of 57 simulations spanning the $\sigma_8-\Omega_M$ parameter plane.
The DES Y1 peaks measurement used the \textsc{Slics} \cite{HarnoisDeraps2021peaks} simulations, spanning 26 points in the $\sigma_8,\Omega_M,h,w$ parameter space.
Future datasets, such as \textsc{CosmoGrid} \cite{CosmoGrid} will consist of orders of magnitude more simulations at higher resolutions, sampling larger parameter spaces and will include more observables beyond the density fields. 

The usage of these simulation grids comes with two significant challenges. Firstly, the sheer size of these datasets may limit their practical applications, as it will require significant storage resources as well as very fast IO systems, which are of limited availability.
The second challenge is the marginalisation of systematic parameters, such as Intrinsic Galaxy Alignments (see \cite{Kirk2015ia} for review), redshift errors, shear calibration and baryon effects \cite{Schneider2020baryons,Lu2021baryons}.
These effects can be typically added to the mass maps using fast algorithms, without running additional N-body simulations.
However, a joint inference of cosmological and systematic parameters requires exploration of high-dimensional parameter space, which is a difficult task for number of dimensions $\gtrsim 10$.
Non-iterative approaches would require large number of these augmented simulations; those could be either based on likelihood modelling on a grid \cite{zurcher2021,HarnoisDeraps2021peaks}, or on a simple Approximate Bayesian Computation (ABC) rejection sampling.
Iterative approaches can converge much faster by exploring this space adaptively.
Several such methods have recently been proposed, based on Monte Carlo-based techniques \cite{Jennings2017astroabc}, or on active learning \cite{Alsing2019pydelfi}.
These methods chose new points in an informed way over the entire parameter space, which is difficult to do on a fixed grid of pre-computed simulations.
Moreover, some of the methods used in the inference process rely on derivatives of maps with respect to cosmological parameters \cite{Charnock2018}; these are typically calculated using finite difference methods. 

Generative machine learning models have recently been proposed to address some of these limitations \cite{rodriguez2018fast,original_cGAN,perraudin2019cosmological,Troester2019painting,Guisarma2019neutrino}. 
In these methods, a generator, which is typically a deep neural network, is trained on simulated data to learn how to transform a random input vector into a new, synthetic density mass map. 
This is possible because the information content at a given resolution is limited, and a finite set of simulations is sufficient to learn the distribution of maps to a very good approximation.
Multiple flavours of these approaches have been proposed, based on Generative Adversarial Networks (GAN) \cite{gan}, Variational Auto-Encoders (VAE) \cite{kingma2013auto}, and U-nets \cite{Ronneberger2015unet}, where other observables are repainted onto the dark matter density field \cite{Guisarma2019neutrino,He2019learning}.
In our recent work \cite[hereafter \citetalias{original_cGAN}]{original_cGAN}, we introduced a map-level emulator that generates weak lensing convergence maps, conditioned on the $\sigma_8$ and $\Omega_M$ parameters, for a given source redshift distribution.
The maps generated by the emulator were in very good agreement with the original dataset, both in terms of the map structure (histograms, power spectra, bispectra, Minkowski functionals) and the variability of the maps (compared using covariance matrices of power spectra as a function of cosmology).

These types of emulators can be of important practical use, as they effectively compress the large simulation grid and can generate hundreds of new maps extremely fast, i.e. in a matter of seconds.
Thus, the map-level emulators address several bottleneck problems: access to non-Gaussian information, simulations dataset size and portability, and the joint, consistent prediction of the map signal and its variation.
Moreover, as they can interpolate between the grid points, they can be useful for aforementioned iterative exploration of the joint cosmological and systematic parameter space, enabling approaches such as ABC or active learning to take advantage of existing simulation grids, as well as methods requiring differentiable simulators.
In this paper, we bring this idea even closer to practical use; we present an emulator of tomographic convergence maps corresponding to the KiDS-1000 survey \cite{kids1000_redshift}, which we call \emuname.
The noise-free maps are generated simultaneously and consistently in five redshift bins, as a function of $\sigma_8$ and $\Omega_M$ parameters, on the sphere, using the \texttt{HEALPix} format.
The emulator uses a conditional GAN architecture with \deepsphere\ \cite{deepsphere} networks, and is trained on the \textsc{DarkGridV1} simulation set \cite{dominik_DES}.
We extensively test the generated sets of maps to the original dataset in terms of cosmological summary statistics and computer vision metrics.
We publish the model and dataset on \texttt{TensorFlow Hub}\footnote{\url{https://tfhub.dev/cosmo-group-ethz/models/kids-cgan/1}}.

The paper is structured as follows:
In Section~\ref{sec:cgan}, we present the method and the training procedure.
Section~\ref{sec:data} contains the description of the dataset.
In Section~\ref{sec:metric}, we describe the metrics that we use to assess the quality of the generator.
We present the results in Section~\ref{sec:results} and conclude in Section~\ref{sec:conclusions}.


\section{Conditional mass map emulator}
\label{sec:cgan}

We implement the mass map emulator using a generative adversarial network (GAN) \cite{gan,Creswell2018ganreview}. GANs consist of two neural networks competing against each other. The concept of adversarial training stems from game theory in a repeated zero-sum game setting. The generative network $G$ produces fake (generated) samples from random noise, while the discriminative network $D$ evaluates both real samples from the training data and the generated ones. The training objective of this generator $G$ is to `deceive' the discriminator $D$ into believing the generated samples are real, whereas $D$ aims to distinguish the real samples from the generated ones to the best of its ability. The model alternates between training $D$ and $G$, and the networks' trainable parameters (weights and biases) are updated through standard backpropagation.
Mathematically, the game played between $G$ and $D$ corresponds to a minimax objective.

Training a GAN model is inherently challenging due to its competitive nature, where the gain of one player is the loss of the other. In practice, the common problems often encountered by GANs include (i) non-convergence \cite{non_convergence_nie2020towards}, as reaching a Nash equilibrium in a two-player non-cooperative game is difficult and may lead to unstable training behaviours such as oscillating losses between $G$ and $D$; (ii) vanishing gradients \cite{wgan}, e.g. when the discriminator significantly outperforms the generator, the generator can no longer learn since the discriminator would always reject the generated samples; and (iii) mode collapse \cite{wgan_gp}, where the generator always produces the same set of outputs as $D$ fails to punish $G$.

To mitigate these problems, various types of regularisation methods \cite{Roth2017, zhang2020consistency} were utilised in recent works concerning GANs. Here, we choose the Wasserstein GAN \cite{wgan} with the implementation of gradient penalty \cite{wgan_gp} (WGAN-GP) as the basis of our generative framework. 
Additionally, we use the conditional version of a GAN \cite{Mirza2014cgan}, meaning both the generator $G$ and discriminator $D$ will be conditioned on some external information $\boldsymbol{y}$. Hence, the objective function of our conditional WGAN-GP is
\begin{equation}
\label{eq:wgan-gp_minimax}
    \min_G \max_D \mathop{{}\mathbb{E}}_{(\boldsymbol{x},\boldsymbol{y})\sim \mathbb{P}_r}[D(\boldsymbol{x},\boldsymbol{y})] -
    \mathop{{}\mathbb{E}}_{\boldsymbol{z}\sim \mathbb{P}_z, \boldsymbol{y}\sim \mathbb{P}_y }[D(G(\boldsymbol{z},\boldsymbol{y}))] - \lambda \mathop{{}\mathbb{E}}_{(\boldsymbol{x},\boldsymbol{y})\sim \mathbb{P}_r \cup \mathbb{P}_g}[(\norm{\nabla_{\boldsymbol{x}}D(\boldsymbol{x},\boldsymbol{y})}_2 - 1)^2],
\end{equation}
where $\mathbb{E}$ is the expectation value, $\mathbb{P}_r$ and $\mathbb{P}_g$ are the real data and generated data distributions respectively. The inputs to the discriminator $\boldsymbol{x}$ are sampled from the real data distribution, and the inputs to the generator $\boldsymbol{z}$ sampled from some noise (latent) distribution $\mathbb{P}_z$, typically a Gaussian. The $\lambda > 0$ is the penalty coefficient of the so-called gradient penalty (GP) term, which directly constrains the gradient norm of the discriminator to be close to unity, in order to enforce the 1-Lipschitz constraint necessary to optimise the Wasserstein distance \cite{wgan_gp}.
Therefore, the role of the discriminator has effectively become that of a critic, as $D$ now outputs a scalar score, which represents how `realistic' the input samples are on average, rather than a probability of whether they are real or not. Empirically, this loss function is shown to correlate with the quality of the generated samples and the convergence of $G$, unlike the case for traditional GAN, and WGAN generally improves stability of training and prevents mode collapse \cite{wgan}.

\begin{figure}
    \centering
    \includegraphics[width=1\textwidth]{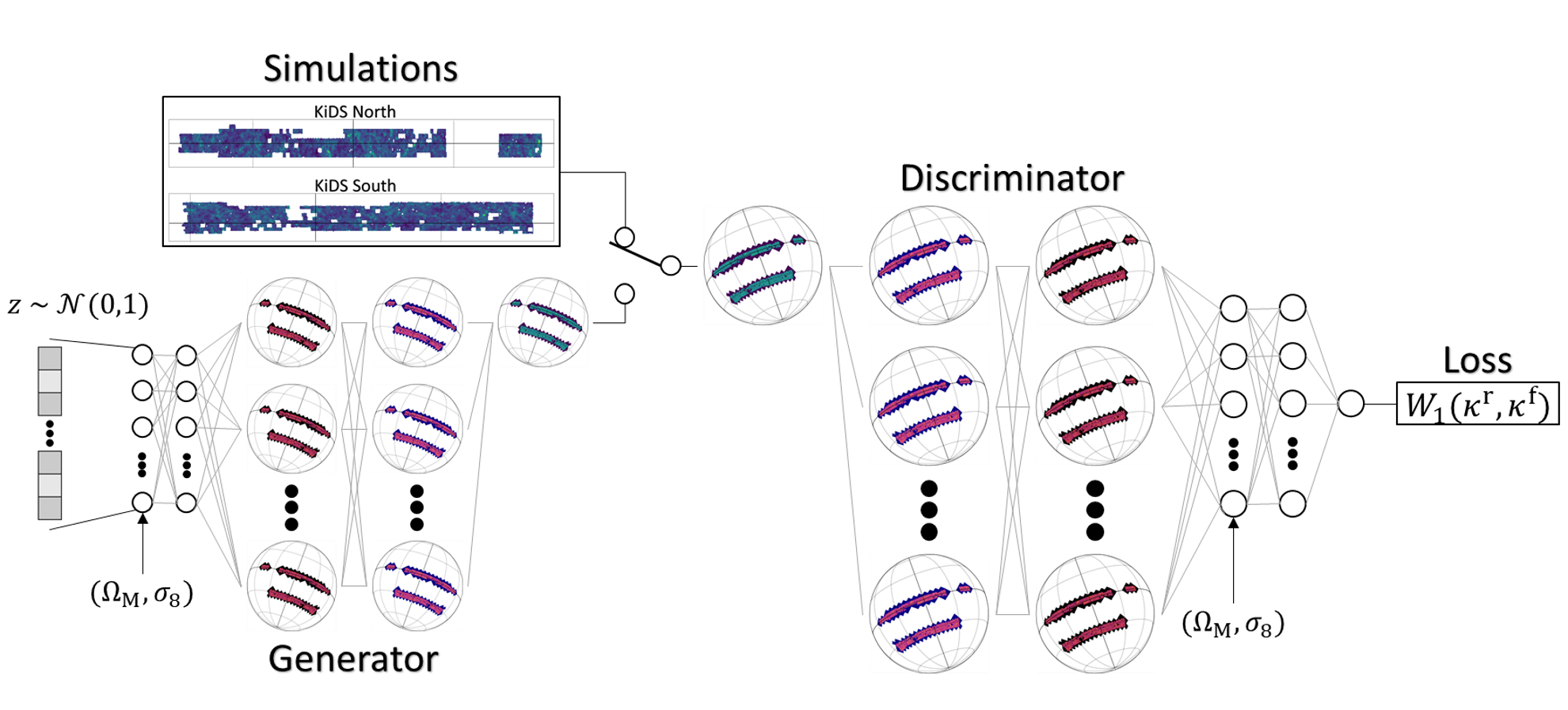}
    \caption{Schematic for the architecture of the conditional GAN used in this work. 
    The random input vector $z$ for the generator, with the length of 584, follows an uncorrelated Gaussian distribution. 
    The conditional variables $\Omega_M$ and $\sigma_8$ are concatenated to the first layer of the generative network. 
    The initial fully-connected layer is followed by a set of spherical feature maps (see Table~\ref{table:archi} for architecture details), represented using a graph network in the \deepsphere\ model.
    The discriminator follows the transposed architecture of the generator.
    The total number of trainable parameters is $\sim2\times$5 million, with the output of the generator being 5 tomographic KiDS-1000-like maps.
    }
    \label{fig:schematic}
\end{figure}

\subsection{Conditioning on cosmological parameters} \label{sec:conditioning}
As our aim is to produce cosmology-dependent, tomographic mass maps, we use conditional GAN (cGAN) \cite{Mirza2014cgan} as mentioned in Equation \ref{eq:wgan-gp_minimax}, where the generated samples can be varied according to a set of variables.
In this paper, the external information $\boldsymbol{y}$ on which the GAN model is conditioned is $\Omega_M$, the total matter density of the Universe, and $\sigma_8$, the linear theory amplitude of mass density fluctuations at the $8h^{-1}$ Mpc scale, also known as clustering strength. The weak lensing (WL) convergence maps are most sensitive to these two cosmological parameters, and those parameters are the only ones that can be tightly constrained using current or ensuing data (see \cite{Refregier2003review,Kilbinger2015review} for reviews).
In practice, many ways exist to perform such conditioning (see \cite{heusel2018gans, reed2016generative, odena2017conditional} for examples); in \citetalias{original_cGAN}, it was done by adapting the distribution of the latent vector $\boldsymbol{z}$ according to the conditioning parameter $\boldsymbol{y}$ using a function $\hat{\boldsymbol{z}} = f(\boldsymbol{z},\boldsymbol{y})$ that scales the norm of $\boldsymbol{z}$ according to $\boldsymbol{y}$. However, we find such scaling method to be ineffective in our model, and thus we opt for the straightforward way of directly concatenating the parameters into both the discriminator and generator as additional input layer. The proposed model is sketched in Figure \ref{fig:schematic}.

\subsection{Spherical GAN architecture} \label{sec:spherical_gan}

The success of convolutional neural networks (CNNs) in the regular Euclidean domains \cite{Bronstein2017euclidean} makes them increasingly popular in cosmology. However, because cosmological data, such as the CMB \cite{adam2016planck}, galaxy number counts \cite{abolfathi2017sdssDR14}, and WL convergence/shear measurements \cite{Chang2018massmap}, often come as spherical maps, efforts have been pursued to create CNNs on the sphere. One natural approach would be to perform grid discretisation on the sphere and apply 2D convolution directly \cite{su2017learning}, while another approach would be to utilise spherical harmonics to perform proper convolutions \cite{cohen2018spherical}. Both methods have their limitations, as the former lacks rotation equivariance, while the latter lacks computational efficiency \cite{deepsphere}.

To achieve the delicate balance between these two properties, the \deepsphere\ \cite{deepsphere} architecture was developed, where the spherical CNN is constructed using graph-based representation of the sphere and tailored to the pixelisation scheme of \texttt{HEALPix} \cite{Gorski_2005}, which is a popular sampling method used in cosmology and astrophysics. By leveraging graph convolutions (GCs) and exploiting the hierarchical nature of \texttt{HEALPix} sampling, this algorithm retains the two aforementioned properties while allowing the analysis of partial sky observations. It has been shown experimentally that the performances of models using \deepsphere\ were superior or equal to the 2D CNN and support-vector machine (SVM) counterparts in classification tasks, especially for high noise levels in input data and for data covering only parts of the sphere \cite{deepsphere}. 
For more information about \deepsphere\ and graph convolutions, see \cite{deepsphere, defferrard2019deepsphere, defferrard2020deepsphere} and the code\footnote{\url{https://github.com/deepsphere/deepsphere-cosmo-tf2}}.
In our work, we extend the \deepsphere\ and create deconvolutional layers to create the GAN architecture.


\section{Training Data}
\label{sec:data}

\subsection{KiDS-1000 Data}

In this work, we create a conditional GAN to generate mass maps with properties of the fourth data release \cite{abolfathi2017sdssDR14} of the Kilo-Degree Survey\footnote{\url{http://kids.strw.leidenuniv.nl/}} (KiDS) which is a public survey by the European Southern Observatory (ESO). The fourth data release contains roughly $1000$ deg$^2$ of images, hence the name KiDS-1000. KiDS is observing in four bands (\textit{ugri}) using the OmegaCAM CCD mosaic camera mounted at the Cassegrain focus of the VLT Survey Telescope (VST), a combination that was designed to have a well-behaved and almost round point spread function (PSF), as well as a small camera shear. This makes it ideal for weak lensing measurements. In combination with its partner survey, VIKING (VISTA Kilo-degree INfrared Galaxy survey~\cite{viking_2013}), the number of observed optical and near-infrared bands of the galaxies increases to nine (\textit{ugriZYJHK$_s$}), leading to greatly improved photometric redshift estimates. In this work, we only make use of the galaxy positions to create the survey footprint and the redshift distributions of the cosmic shear analysis \cite{kids1000_shear,kids1000_method} to generate mock convergence maps for the training of the conditional GAN.

\subsection{Simulations}

\begin{figure}
    \centering
    \includegraphics[width=1\textwidth]{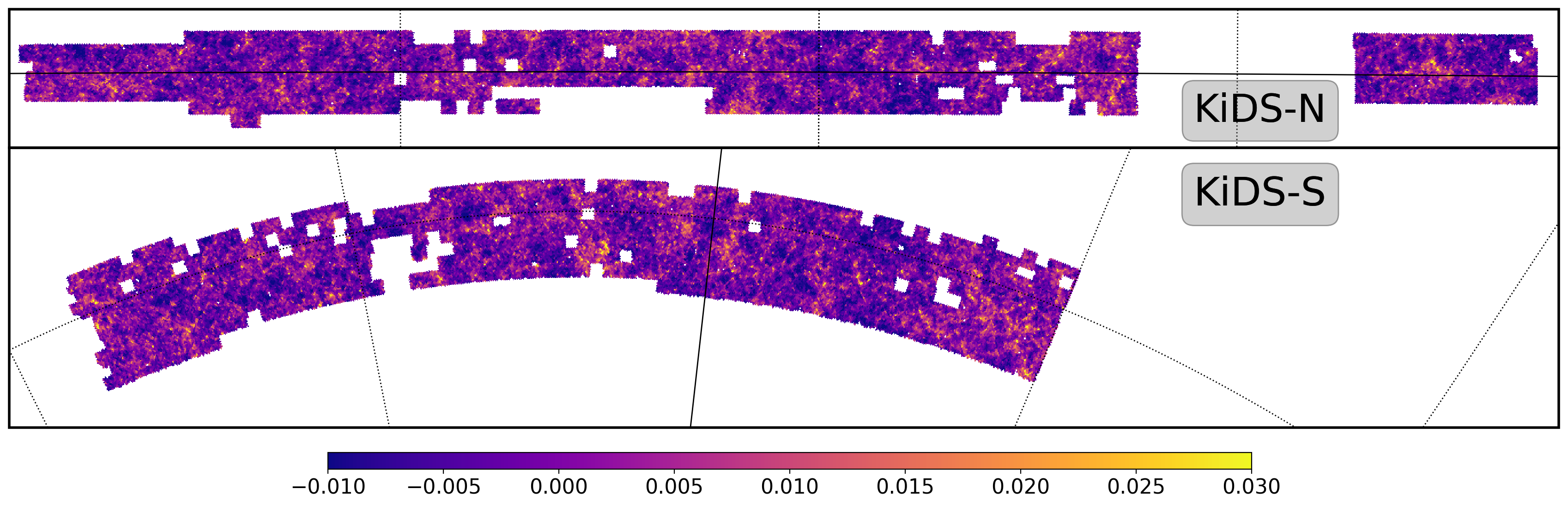}
    \caption{One of the 5 tomographic mass maps ($z$-bin $=5$ here) of a randomly chosen mock survey from the N-body simulations with the cut-outs of the KiDS-1000 footprint, which consists of 3 disconnected regions (the relative distances between these regions are shortened here for visualisation purposes). The colour intensity corresponds to the value of convergence~$\kappa$. The top two patches are from the northern KiDS field and the bottom patch from the southern field \cite{kids1000_method}. The padded area is not shown here. The graticule is of intervals of 60{\textdegree} between meridians and 30{\textdegree} between parallels.}
    \label{fig:real_sample}
\end{figure}

\begin{figure}
    \centering
    \includegraphics[width=.5\textwidth]{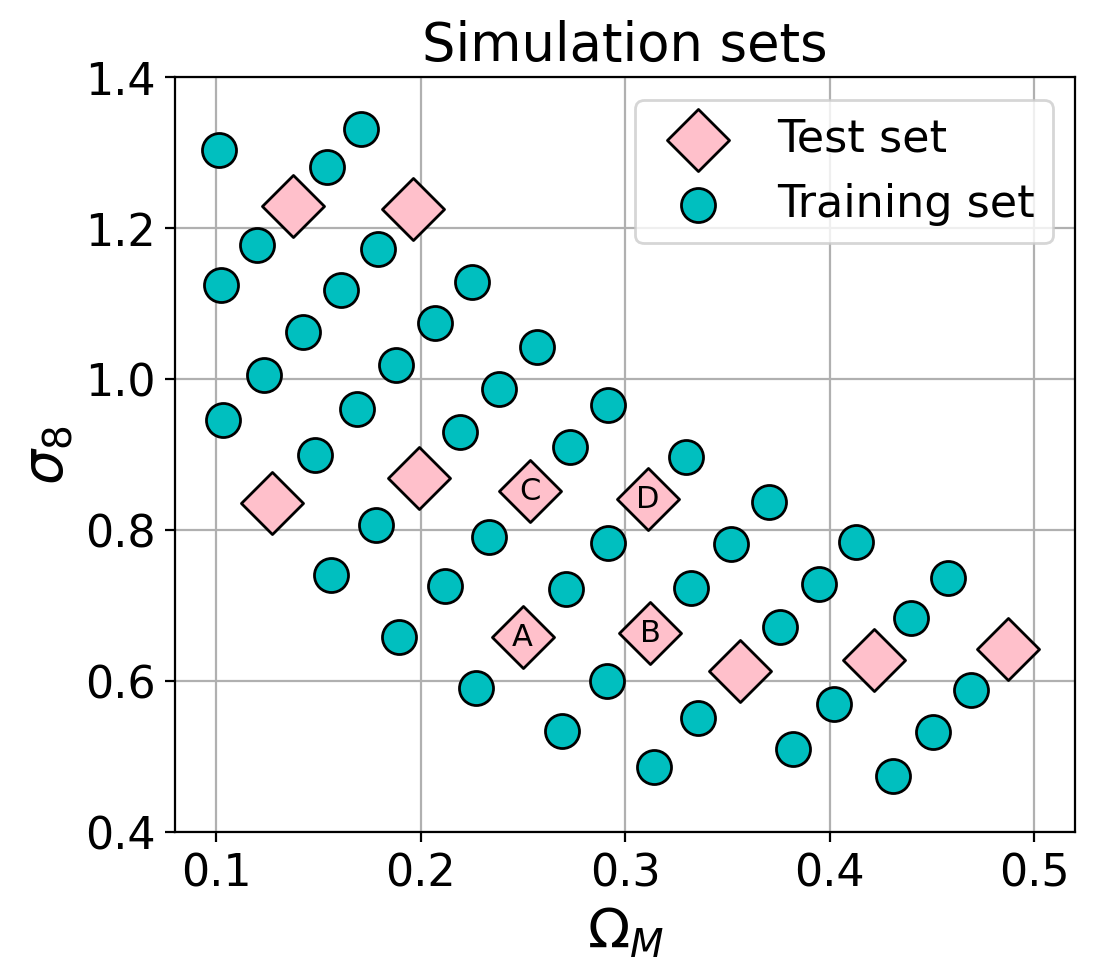}
    \caption{Cosmological parameter grid from \citetalias{original_cGAN}, also used in this work. The diamonds and the circles represent the 11 test sets and 46 training sets respectively. The cosmologies marked A, B, C and D are investigated in more detail in Section \ref{sec:results}.}
    \label{fig:train_test_set}
\end{figure}

The training of a conditional GAN requires a significant amount of data. We generate the maps using the N-body simulation suite from \cite{zurcher2021,dominik_DES}. It is created using the full-tree N-body code \textsc{PkdGrav3}~\cite{pkdgrav3} and provides 5 independent simulations for 57 different cosmological parameters in the $\Omega_M - \sigma_8$ plane, plus an additional 50 distinct simulations of the fiducial cosmology ($\Omega_M = 0.26$, $\sigma_8 = 0.84$). The grid is equivalent to the one used in \citetalias{original_cGAN} and shown in Figure~\ref{fig:train_test_set}. Each simulation is performed using $768^3$ dark matter particles in a box with a side length of 900 Mpc/$h$ and produces a past lightcone of $\sim80$ discrete shells with a \texttt{HEALPix} resolution of $N_{\textrm{side}} = 2048$, spanning a redshift range $z$ from 0 to 3. The grid contains flat $\Lambda$CDM cosmologies, with all remaining parameters fixed to the ($\lambda$CDM,TT,TE,EE+lowE+lensing) results from Planck 2018 \cite{planck2018}. More details about the simulations can be found in \cite{zurcher2021}.

For each simulation, we project the discrete particle shells to full-sky convergence maps using the \textsc{UFalcon} code~\cite{sgier1,sgier2} with the redshift distributions of the 5 tomographic bins used in the KiDS-1000 cosmic shear analysis \cite{kids1000_shear,kids1000_method}. The first 4 bins are spaced by $\Delta z = 0.2$ in the range of $0.1 < z \leq 0.9$ while the 5th bin covers $0.9 < z \leq 1.2$ \cite{kids1000_redshift}. To increase the amount of training data, we randomly rotate the full sky maps 200 times for the grid and 50 times for the fiducial simulations. Each rotation is performed with a resolution of $N_{\textrm{side}} = 2048$. Afterwards we downsample the maps to the final $N_{\textrm{side}} = 512$ to mitigate possible artefacts introduced by the rotation of the finite pixels. Finally, we cut out the KiDS-1000 footprint using only pixels that contain at least one galaxy of the catalogue \cite{kids1000_cata} and subtract the mean of the maps for each redshift bin. This procedure results in 1000 mock surveys (5 simulations $\times$ 200 rotations) for each of the 57 cosmologies for the training and 2500 mocks (50 simulations $\times$ 50 rotations) of the fiducial cosmology. See Figure~\ref{fig:real_sample} for visualisation of tomographic maps used in this work.

By construction of \deepsphere, appropriate padding to the patch is necessary to ensure the maps/latent vector can be properly downsampled/upsampled when passing through the discriminator/generator to the wanted $N_{\textrm{side}}$ during the training cycles. Here, the number of pixels for each tomographic map is around 74000 at $N_{\textrm{side}} = 512$. Due to the distribution of source galaxies in the 5 tomographic bins \cite{kids1000_method} in the KiDS-1000 survey, each tomographic map has a slightly different amount of pixels: In this work, $N_\textrm{pixels}=74057,74375,74538,74449,74406$ for $z$-bins 1 to 5 respectively.
In our model, we desire to reduce the $N_{\textrm{side}}$ of the maps to 32, which corresponds to passing through four standard \deepsphere\ strided convolutional layers in the discriminator. This implies the number of pixels must be divisible by $16^2 \times 4$. By padding minimally, each map now consists of 149504 pixels exactly, which corresponded to 584 `superpixels' when downsampled to $N_{\textrm{side}} = 32$.

As the results of \citetalias{original_cGAN} serve as a benchmark for our model, the dataset is split into a training and test set the same way as \citetalias{original_cGAN}, shown in Figure~\ref{fig:train_test_set}. Only 46 out of the 57 cosmologies would contribute to the training sets, while the remaining 11 cosmologies would serve as the test sets, i.e. they are not fed into the network during the training process. This way, we enable the evaluation of the capability of the model to interpolate to unseen cosmologies. Each of these training sets and test sets consists of 1000 mock surveys, each with 5 tomographic maps.
The original \citetalias{original_cGAN} paper performed detailed analysis of the summary statistics for the cosmologies from the test set marked with letters A, B, C, and D. However, as our data now consist of tomographic maps, the abundance of information would make it hard to illustrate the quantitative comparison of the original N-body-simulated maps and the GAN-generated maps if we were to show all 4 cosmologies in detail. Therefore, for the sake of brevity, in Section~\ref{sec:results} we show the detailed summary statistics and the corresponding figures only for the best and worst models in the test set (A and D respectively). Nevertheless, the summary statistics of all models, including B and C, are evaluated quantitatively.


\section{Quantitative comparison metrics}
\label{sec:metric}

One of the main challenges of training GANs is the lack of standard metrics to evaluate the performance due to the competitive nature of $D$ and $G$ encapsulated in Equation~\ref{eq:wgan-gp_minimax}, which is not the case with traditional neural networks, where a low validation loss directly corresponds to a better model. This is less important in natural image processing, in which case the sole purpose of the generator is to generate photo-realistic images from a human's perspective.
However, to perform a quantitative assessment of the results for cosmological maps, we require well-defined metrics and summary statistics, rather than solely relying on visual inspection.
Multiple summary statistics of WL convergence maps have already been studied \cite{zurcher2021, mandelbaum2018weak}, and thus we can utilise them to evaluate the quality of the generated samples in a statistical manner. 
We expand the set of metrics in \citetalias{original_cGAN}, with the full list as follows:
\renewcommand{\theenumi}{\roman{enumi}}%
\begin{enumerate}
    \item the angular power spectra $C_\ell$, including both the auto spectrum of a singular tomographic map, and the cross spectrum between two tomographic maps,
    \item the pixel distribution of the convergence maps $N_{\textrm{pixels}}$, compared using histograms and the Hellinger distance,
    \item the peak distribution of the convergence maps $N_{\textrm{peaks}}$, also compared using histograms and the Hellinger distance,
    \item the Pearson’s correlation matrices $R_{\ell\ell'}$ between the $C_\ell$ of the tomographic maps at different cosmologies, and
    \item the Structural Similarity (SSIM) Index \cite{wang2004ssim}, an image comparison metric commonly used in computer vision.
\end{enumerate}
\noindent
Originally, the Wasserstein-1 distance $W_1$ was used for the pixel/peak distribution comparisons in \citetalias{original_cGAN}; $W_1$ has become a standard metric for machine learning and computer visions in recent years \cite{Frogner2015learning, Kolouri2019Wasserstein}, as it was shown to provide a useful tool to quantify the geometric discrepancy between two distributions \cite{Kolouri2015Wasserstein}.
However, since we have chosen a Wasserstein GAN as the basis of our generative framework, it would be useful to have a distance other than $W_1$ as a metric to gauge the overall performance of the GAN results. Therefore, we opted for another well-defined distance, called Hellinger distance \cite{Hellinger1909}, which has found numerous applications\footnote{More specifically, the Bhattacharyya coefficient $BC$ is more common but it is closely related to Hellinger distance: $H = \sqrt{1-BC}$.} in pattern recognition and machine learning \cite{Bhattacharyya2006, Bhattacharyya2003, Bhattacharyya2020}. See Section~\ref{sec:evaluation} for more details regarding the Hellinger distance.

In addition to the usual non-tomographic summary statistics, since our data consist of multiple tomographic maps, we can also utilise tomographic statistics as part of quantitative assessment of the cGAN. Namely, the cross power spectra, 2D pixel/peak histograms, and cross-correlation matrices would complement their non-tomographic counterparts in the following sections.

To validate if the cGAN emulator can be used for creating cosmological constraints, we perform a mock cosmological analysis with several summary statistics of the maps: angular power spectrum $C_\ell$, peak statistics and convergence pixel distribution.
Moreover, we demonstrate that a Likelihood-Free Inference (LFI) approach, \textsc{PyDelfi} \cite{Alsing2019pydelfi}, can be effectively deployed. For the detailed descriptions of the aforementioned metrics and cosmological constraints, see Appendix~\ref{app:metrics_description}.

\subsection{Performance evaluation of the cGAN emulator} \label{sec:evaluation}

Therefore, the metrics for evaluating the performance of the cGAN include human inspection, the aforementioned cosmological summary statistics, the SSIM Index, and the cosmological constraints. Note that visual evaluation was just as important as the other summary statistics, since it is possible for the generator to produce `nonsensical' samples, e.g. from mode collapse, which could still be consistent with the statistics of the real data.
We then perform the evaluation of the mass map emulator by computing the summary statistics of the GAN-generated convergence maps with the cosmological parameters from the test sets, which are marked by the diamonds in Figure \ref{fig:train_test_set}.

Then, we compare these generated statistics to the statistics from the corresponding N-body-simulated maps.
To do this, $N=1000$ generated samples are created after the training process of the cGAN. We then compute the summary statistics of the generated data to be compared to that of the $M=1000$ real samples. 
Now, given that we have real maps $\{x_{r,i}\}$ for $i=1,...,M$ and generated maps $\{x_{g,i}\}$ for $i=1,...,N$, we can compute the mean $\mu$ and standard deviation $\sigma$ of real/generated statistics by averaging all the N-body/GAN realisations.
Originally, \citetalias{original_cGAN} compared the histograms between the N-body and GAN maps pixel/peak distributions $x$ using the normalised Wasserstein-1 distance (see Appendix~\ref{app:peak_hist} for the definition). Here, we use the aforementioned Hellinger distance instead, defined as
\begin{equation}
\label{eq:hellinger}
    H(\hat{x}_{\textrm{N-body}}, \hat{x}_{\textrm{GAN}}) = \frac{1}{\sqrt{2}} \left( \sum^{k}_{i=1}(\sqrt{r_i} -\sqrt{g_i})^2 \right)^\frac{1}{2},
\end{equation}
where $\hat{x}$ denotes the normalised pixel/peak distribution, and $r_i, g_i$ comprise the discrete pixel/peak distributions (histograms) of the N-body and GAN maps respectively, i.e. $\hat{x}_{\textrm{N-body}} = (r_1, ..., r_k)$ and $\hat{x}_{\textrm{GAN}} = (g_1, ..., g_k)$, with $k$ being the number of histogram bins. 
This quantity is the distribution analogue of the Euclidean distance. 
Additionally, using the properties of the distributions, it is easy to show that
\begin{equation}
\label{eq:hellinger_simple}
    H(\hat{x}_{\textrm{N-body}}, \hat{x}_{\textrm{GAN}}) = \left( 1-\sum^{k}_{i=1}\sqrt{r_i g_i} \right)^{\frac{1}{2}}.
\end{equation}
Now, one can see that the Hellinger distance $H$ is bounded between 0 and 1, with the former implying the two distributions are identical. Therefore, this makes the metric easy to interpret as an ideal GAN would generate a pixel/peak distribution such that $H$ is as close as 0 (similar to the interpretation of $W_1$ where a value closer to 0 generally implies a better GAN model). Additionally, since $H$ is scale-independent, it can be used to quantitatively compare different GAN models.
To give some intuition about the Hellinger distance, we calculated that, for a univariate Gaussian, a 1$\sigma$ shift in the mean corresponds to $H \approx 0.34$ and doubling the $\sigma$ to $H \approx 0.32$, whereas a $5\%$ shift in the mean corresponds to $H \approx 0.017$, and $5\%$ difference in $\sigma$ to $H \approx 0.024$.

For the auto and cross power spectra comparison, the agreement is simply quantified by the fractional difference between the real and generated statistics:
\begin{equation}
\label{eq:psfracdiff}
 f_{C_\ell}= \left\langle \frac{C_{\ell}^{\textrm{GAN}}-C_{\ell}^{\textrm{N-body}}}{C_{\ell}^{\textrm{N-body}}} \right\rangle_{\ell \in [100,1000].} 
\end{equation}
We compare the correlation matrices using Frobenius norms $\norm{R}_\textrm{F} \equiv (\sum_{i,j}\abs{r_{ij}}^2)^\frac{1}{2}$, by taking their fractional difference: 
\begin{equation} 
\label{eq:frobenius}
    f_R=\frac{\norm{R_{\textrm{GAN}}}_\textrm{F}-\norm{R_{\textrm{N-body}}}_\textrm{F}}{\norm{R_{\textrm{N-body}}}_\textrm{F}}.
\end{equation}
To quantify the agreement in the SSIM metric (see Equation~\ref{eq:ssim} in Appendix \ref{app:ssim}), we consider the significance of the difference in SSIM measurements of 500 pairs of N-body/GAN maps:
\begin{equation}
\label{eq:sig_diff}
    s_\textrm{SSIM}=\frac{\langle\textrm{SSIM}_\textrm{GAN}\rangle-\langle\textrm{SSIM}_\textrm{N-body}\rangle}{(\sigma[\textrm{SSIM}_\textrm{GAN}]+\sigma[\textrm{SSIM}_\textrm{N-body}])/2},
\end{equation}
where the $\langle\textrm{SSIM}\rangle$ is the mean score, and $\sigma[\textrm{SSIM}]$ is the standard deviation. As mentioned in Section \ref{sec:cgan}, a successful GAN model should preserve the data statistics and produce samples as diverse as the real N-body samples. Therefore, a small $s_\textrm{SSIM}$ (i.e. a good agreement in the SSIM scores) indicates the absence of mode collapse, and a large $s_\textrm{SSIM}$ suggests otherwise.

\subsection{Implementation} \label{sec:implementation}

All codes used in this paper are written in Python 3.8+. Our cGAN model is published on \texttt{TensorFlow Hub}\footnote{\url{https://tfhub.dev/cosmo-group-ethz/models/kids-cgan/1}}. The model is created using \deepsphere\ and implemented in the open-source library \texttt{TensorFlow} \cite{tensorflow}, along with the \texttt{Keras} deep learning API \cite{chollet2015keras}.
The model is trained on the Piz Daint GPU cluster of Swiss National Supercomputer Centre (CSCS).

The angular power spectra $C_\ell$ are calculated directly from the simulated mass maps with no smoothing using \ttvar{SurveyEmulator.power_spectra}, which is a GPU compatible implementation similar to the \texttt{anafast} routine from \texttt{healpy}. See our \texttt{TensorFlow Hub} site for more details.
For the mass map histograms, we simply count the value of every pixel in every map for each cosmology, excluding the zeros in the padded region, and obtained the mean and standard deviation of the histograms. The same procedure is applied to the peak histograms, in addition to searching all pixels with a value larger than its 8 (or 7)\footnote{The 48 pixels at the vertices of the base dodecahedron have only 7 neighbouring pixels \cite{deepsphere}.} neighbour pixels and extracting them.
The correlation matrices are computed using the \texttt{numpy} package.
Our quantitative comparison of the cosmological constraints from the mock analysis is based on calculating the mean and standard deviation of $\sigma_8$ and $\Omega_m$ posteriors and comparing them.
We also calculate the Jensen-Shannon divergence (JSD) between the 2D posteriors.


\begin{figure}
    \centering

    \includegraphics[width=0.8\textwidth]{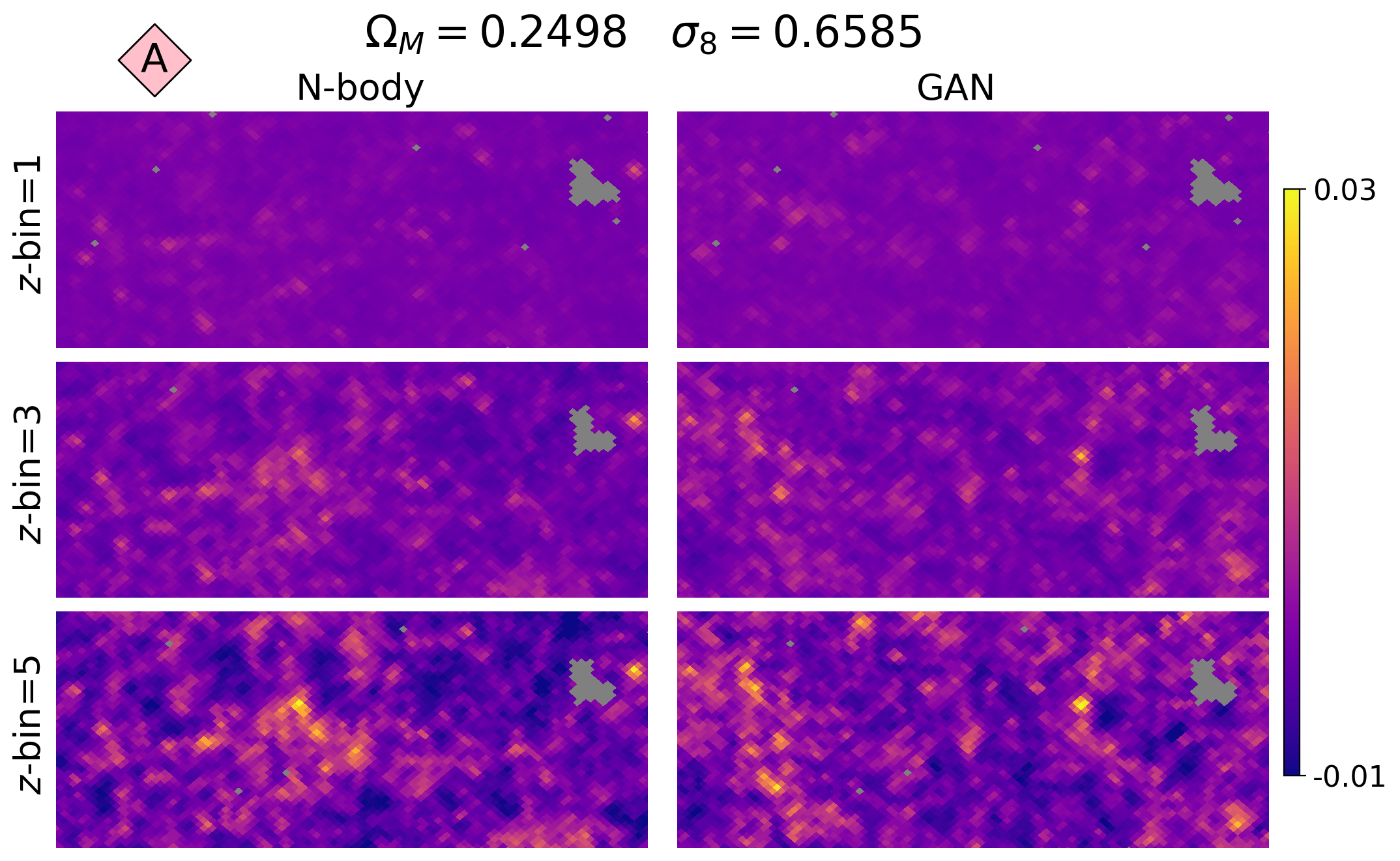}    \includegraphics[width=0.8\textwidth]{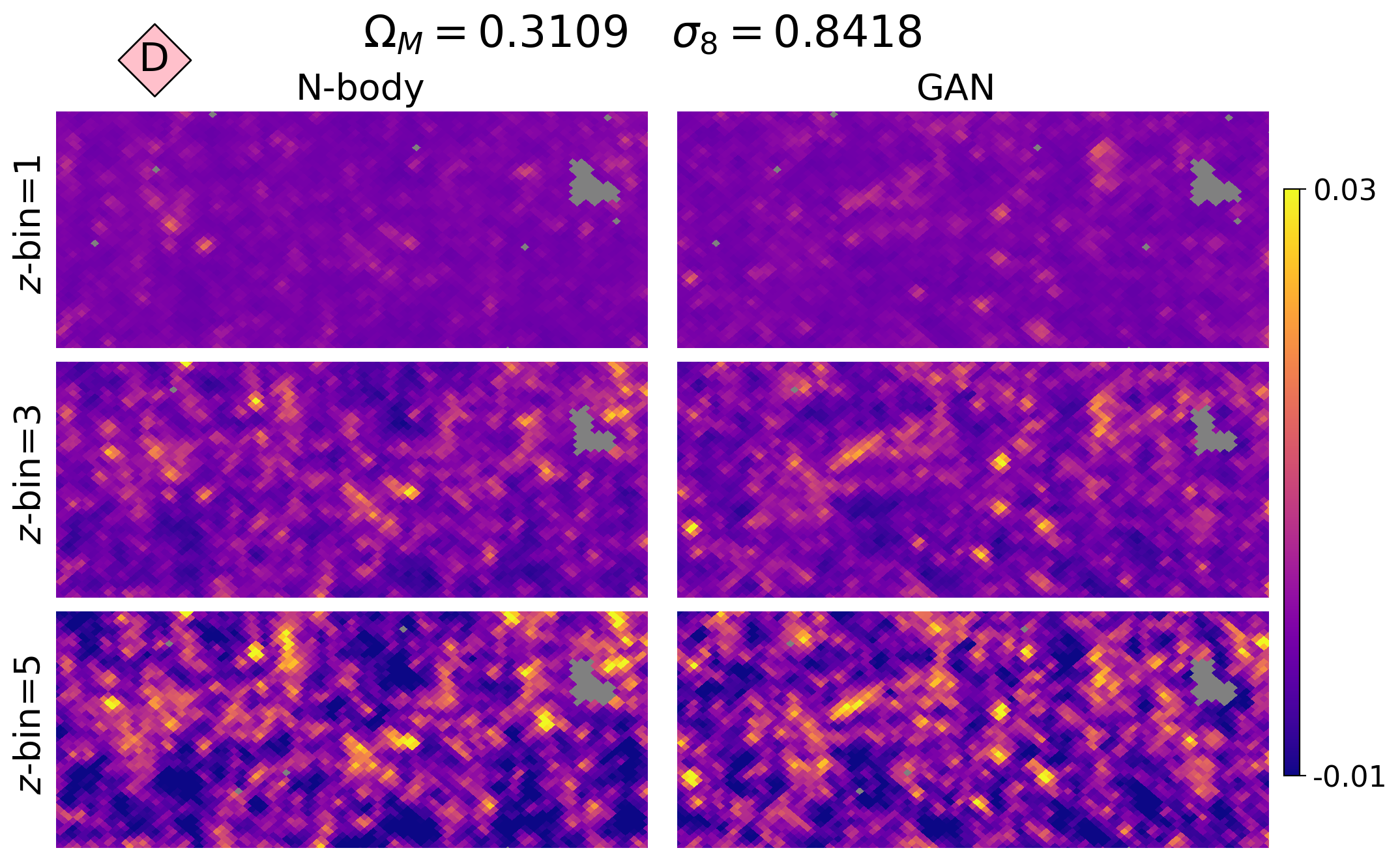}
    \caption{A comparison of the original WL convergence maps from the N-body simulations and the GAN-generated maps for models A and D corresponding to Figure \ref{fig:train_test_set}. The pixel values show the convergence $\kappa$. For clarity, only the tomographic maps with $z$-bin 1, 3, and 5 are shown. The grey regions are empty (i.e. not part of the map).
    The maps are both smoothed with a Gaussian filter of FWHM $=8.1$ arcmins.}
    \label{fig:fig_real_vs_gan}
\end{figure}

\section{Results}
\label{sec:results}

The results shown in the following section are generated by the cGAN with the network architectures shown in Table~\ref{table:archi} in Appendix~\ref{app:architecture}.
In short, we use mirrored architectures for the generator and the discriminator, with 8 and 11 layers (input and output layers included), correspondingly.
The cGAN consists of approximately 5 million trainable parameters for each network.
The model is trained on a single GPU node for around 160 hours on the Piz Daint cluster, which uses Nvidia Tesla P100 GPUs.

\subsection{Visual comparison and latent interpolation}

We present the original and generated convergence maps for test models A and D in Figure \ref{fig:fig_real_vs_gan}. The N-body-simulated and GAN-produced samples are visually indistinguishable across all tomographic bins; the maps generated by $G$ clearly show structures very similar to that from the original maps, and the model can capture the prominent features of mass maps present in the real data, such as overdense and underdense regions.
Moreover, the variance in pixel intensity in the generated maps increases as we look further back in redshift, matching that in the real maps. 
Additionally, the generator is able to produce a wide variety of patches, and we never found any pair of generated samples that looked too similar. This indicates that the GAN model did not collapse into generating only a small set of output samples, i.e. no mode collapse.

\begin{figure}
    \centering
    \includegraphics[width=1\textwidth]{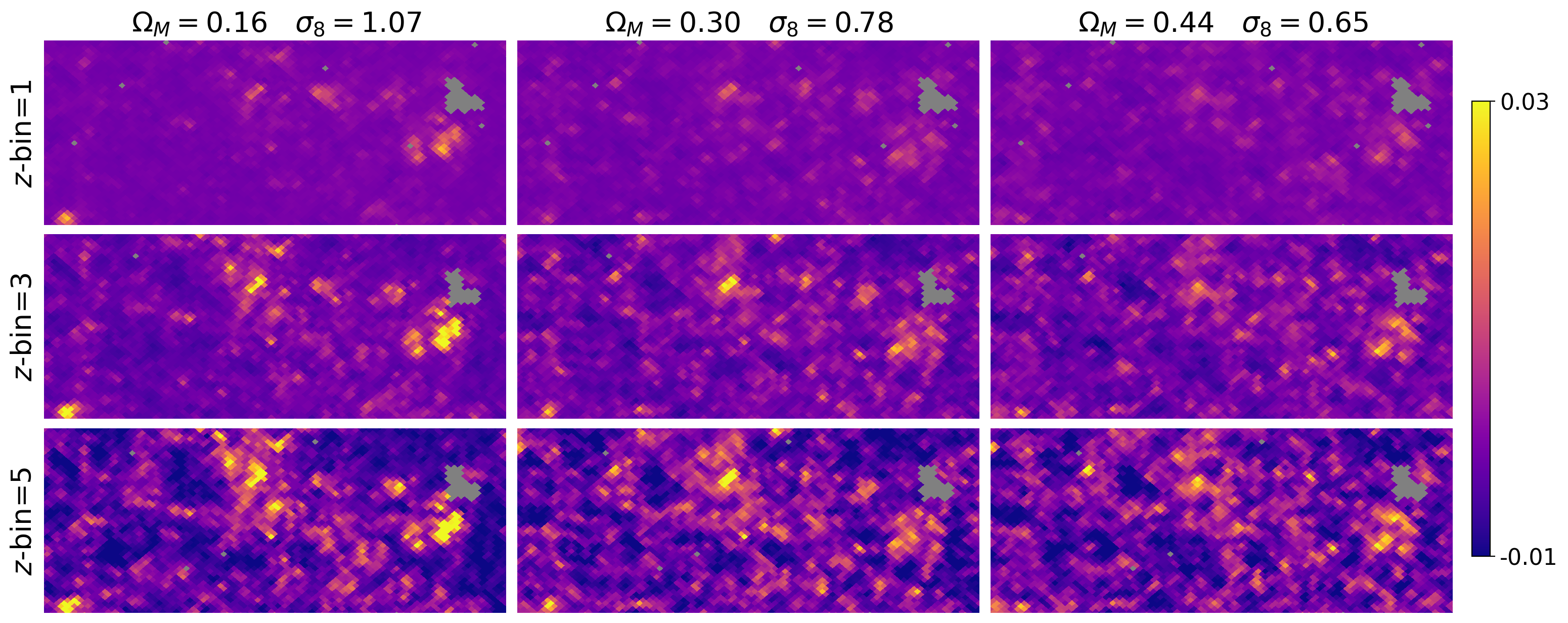}
    \caption{Generated samples from the same random seed, but with varying $\Omega_M$ and $\sigma_8$.
    A fixed value of $S_8=\sigma_8(\Omega_M/0.3)^{0.5}=0.78$ was used, to illustrate how the non-Gaussian features change for models that have very similar power spectra.}
    \label{fig:latent_fixed_S8}
\end{figure}

The interpolating power of the conditional GAN is demonstrated in Figure \ref{fig:latent_fixed_S8}; here, the same latent variable $z$ is fed into the generator along with three combinations of cosmological parameters, chosen to have the same $S_8=0.78$, and thus lie on the $\Omega_M-\sigma_8$ degeneracy line. Such parameter combinations have very similar power spectra.
This illustrates the smooth transitions of the generated maps from one cosmology to another, indicating that the emulator achieves the objective of interpolating between cosmologies.
We also notice the consistency of the transition between the redshift bins.
Here we only show $z$-bins 1, 3, and 5 for brevity, but note that the remaining bins have the same properties.

\subsection{Summary statistics}

\begin{figure}
    \centering
    \includegraphics[width=1\textwidth]{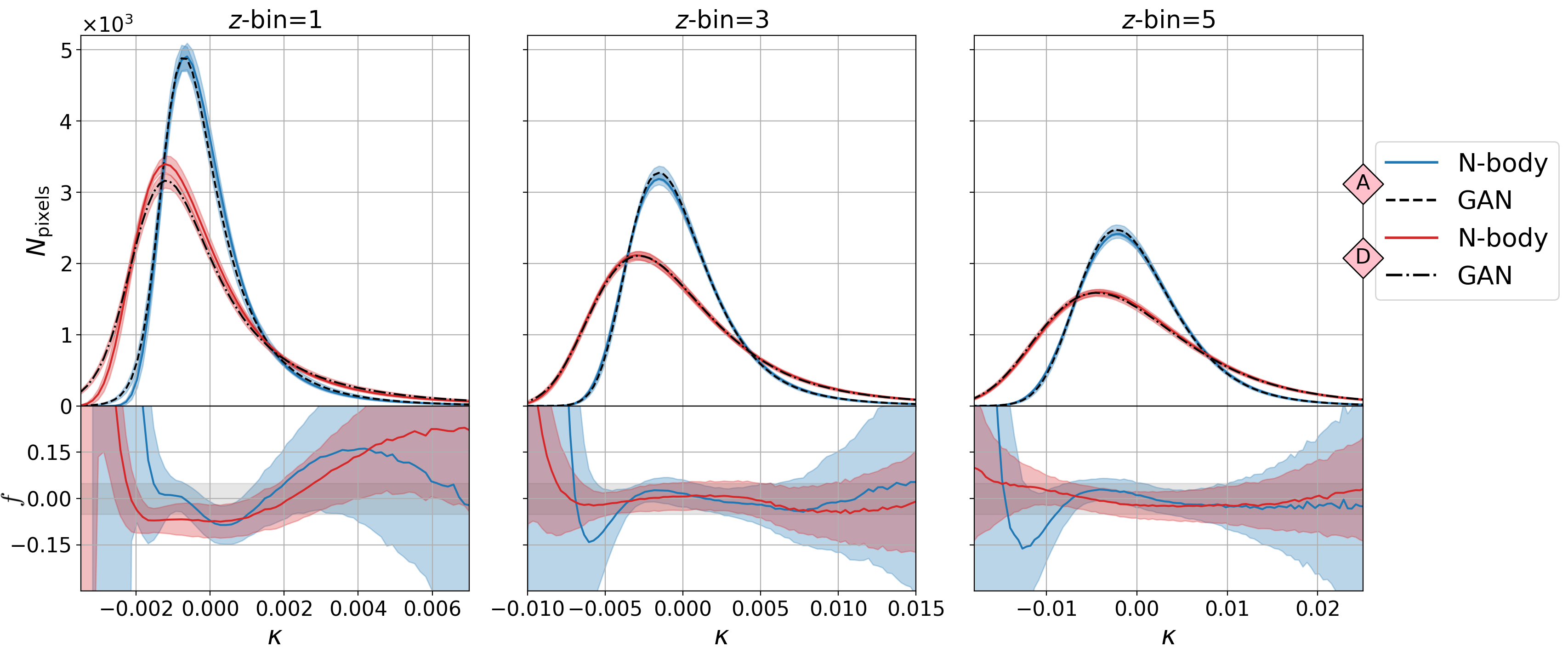}
    \includegraphics[width=1\textwidth]{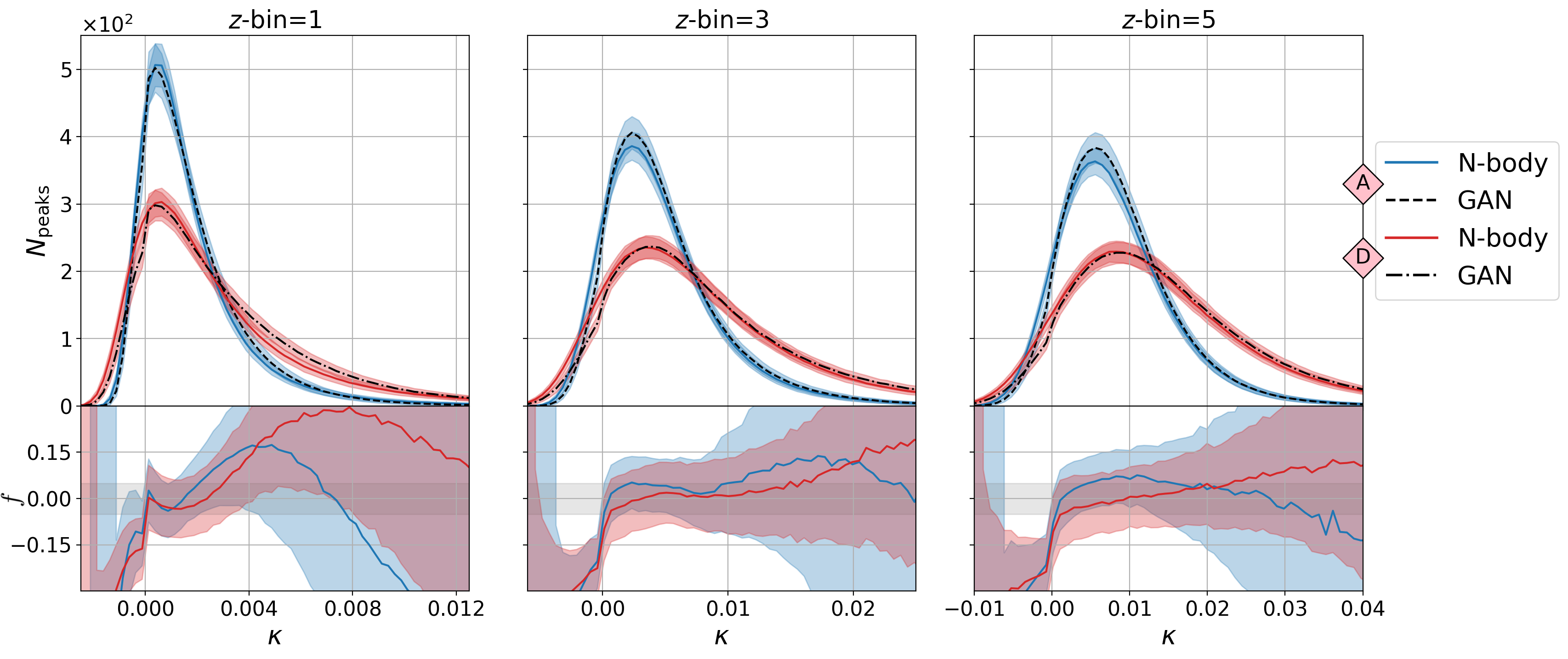}
    \caption{Comparisons of pixel histograms (top) and peak histograms (bottom) between the real N-body-simulated and GAN-generated maps for models A and D for $z$-bins 1, 3 and 5. Here, we choose to show the two models that are the overall {\bf best-performing (A)} and {\bf worst-performing (D)} amongst the test set.
    The real data statistics are computed from the corresponding 1000 test set samples, and the generated counterparts from 1000 GAN samples generated from random seeds. The convergence $\kappa$ was represented by the pixel values. A pixel whose value is greater than that of its 8 neighbouring pixels is defined as a peak. Each coloured band corresponds to one standard deviation of the respective data. The lower panels show the fractional difference $f$ between the median histograms. The coloured bands correspond to their corresponding standard deviations of the ratio distribution of $x_{\textrm{GAN}}/x_{\textrm{N-body}}$.}
    \label{fig:real_vs_gan_pix_n_peak}
\end{figure}

We compute the summary statistics described in Section~\ref{sec:metric} using 1000 real and 1000 generated samples for every cosmology in the test set. The statistics of tomographic maps with $z$-bin 2 and 4 are excluded from the presented plots for the sake of clarity, as they look similar to the neighbouring $z$-bins 1, 3, and 5. Furthermore, though the following figures only show the statistics for models A and D, the results for B, C, and other test set cosmologies are very similar.
Figure \ref{fig:real_vs_gan_pix_n_peak} shows the pixel histograms (top row) and peak histograms (bottom row) of the original maps from N-body simulations (solid) and the GAN-generated maps (dashed), for models A (blue) and D (red) corresponding to Figure \ref{fig:train_test_set}. The solid and dashed lines represent the mean of the 1000 histograms from the real and generated data respectively, and the coloured bands their standard deviation. 
We also calculate the simple fractional difference between the original and generated samples, defined as $f=(x_{\textrm{GAN}}-x_{\textrm{N-body}})/x_{\textrm{N-body}}$, shown in the panel below each histogram. There, the coloured bands correspond to the square root of the variance of ratio distribution, i.e. $\textrm{Var}(x_{\textrm{GAN}}/x_{\textrm{N-body}})=\textrm{E}(x_{\textrm{GAN}}^2) \textrm{E}(1/x_{\textrm{N-body}}^2) - \textrm{E}^2(x_{\textrm{GAN}}) \textrm{E}^2(1/x_{\textrm{N-body}})$, where $E$ is the expectation value. Evidently, the error is much larger at tail-end regions since the bins of the N-body distribution at these $\kappa$ values contain only very few data points.

Overall, we see a good agreement in the pixel histograms. 
In particular for the latter redshift bins of $z$-bins 3 and 5 (also 4, see Appendix \ref{app:additional_plots}), the N-body and GAN histograms agree very well visually, whereas for the first $z$-bin and low $\kappa$ values, the agreement gets considerably worse compared to the other $z$-bin and at high $\kappa$ values. The discrepancy is most apparent at the upper left panel of Figure~\ref{fig:real_vs_gan_pix_n_peak}, where the difference between the means of the N-body and GAN statistics is greater than $1\sigma$ at very negative $\kappa$ values. For example, the upper leftmost panel in Figure~\ref{fig:real_vs_gan_pix_n_peak} shows that the GAN histogram differs significantly from the N-body histogram in our weakest model D at $\kappa \lesssim -0.003$. This signifies the GAN model is weaker in following the statistics in underdense regions of the map. 

The reason for such discrepancy is that GANs tend to focus on learning the bulk of the distribution, as that is where most of the information is located. As such, GANs are known for having difficulties following the tail of the distribution and rare events which contributes to only a small part of the data distribution \cite{Dionelis2021tailGAN}. We expect more advanced GAN architecture such as cycleGAN \cite{Zhu2017cycleGAN} to yield significant improvement in future work. For now, when considering the typical noise of these surveys, the level of accuracy for these statistics is good enough to perform cosmological constraints (see Section \ref{sec:constraints} for more details).
The agreement can be further quantified by computing the Hellinger distance of the pixel values distribution (see Section \ref{sec:metric} for definition): for example, $H_{\textrm{pixels}}$ for model A are $0.0577, 0.0200, 0.0123, 0.0089, 0.0117$ for tomographic bins 1 to 5 respectively. 
Evidently, the Hellinger distance for the first $z$-bin is worse than that of the other $z$-bins, which is consistent with visual inspection of these histograms.
Intuitively, this can be roughly interpreted as an agreement of $\lesssim 5\%$ for $z$-bins 2-5 and around 15\% for $z$-bin 1 (see Section~\ref{sec:metric} for intuition on Hellinger distance).
By averaging over all tomographic bins, the overall Hellinger distances of the pixel distribution $H_{\textrm{pixels}}$ are $0.0221, 0.0199, 0.0348, 0.0337$ for models A, B, C, D respectively. 
Using normalised Wasserstein-1 distance $W_1$ instead of $H$ as a metric also gives very similar result, i.e. the $z$-bin-averaged $W_1$ values of these models are generally small, with the latter $z$-bins having better performance. 

\begin{figure}
    \centering
    \includegraphics[width=1\textwidth]{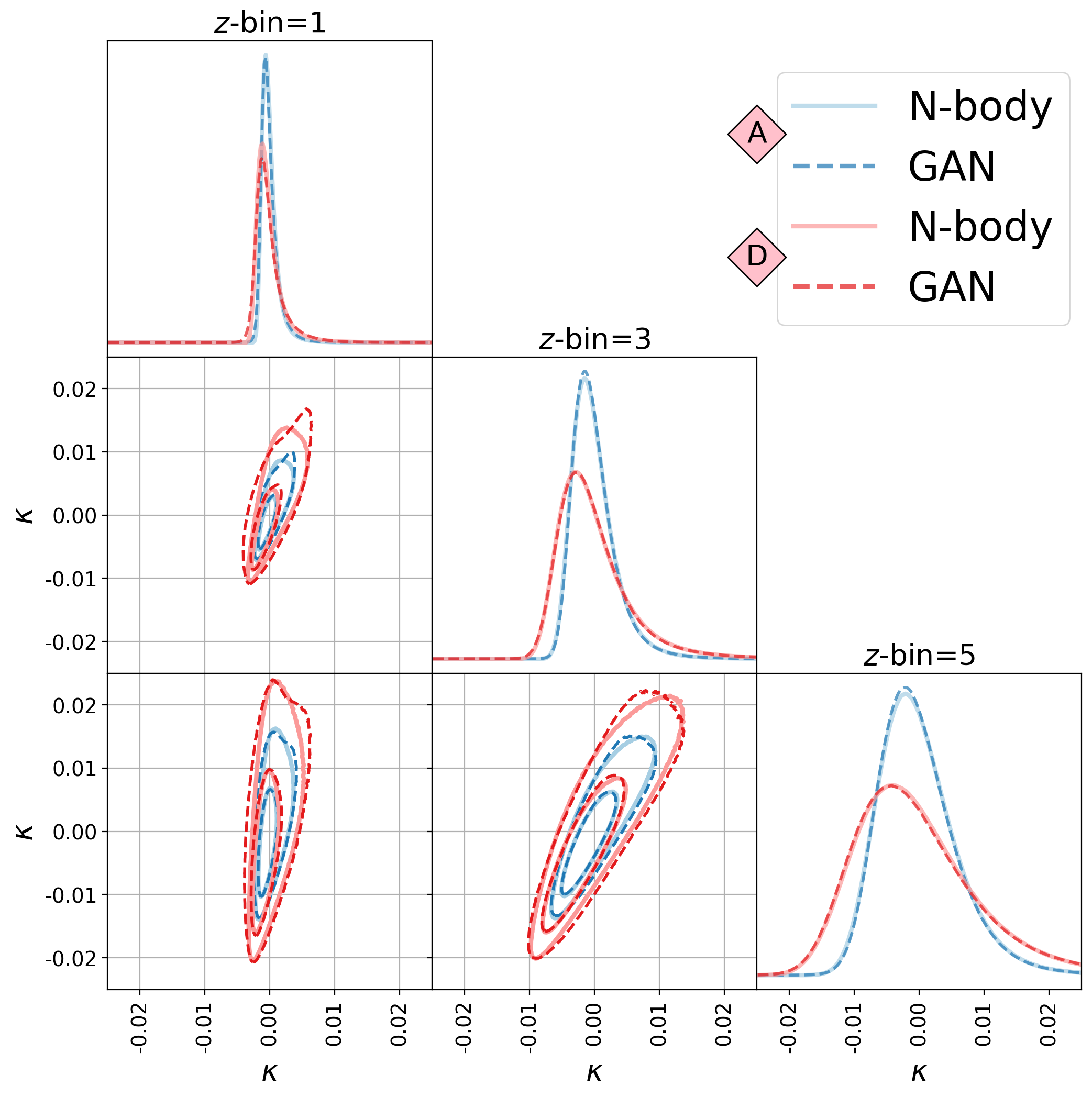}
    \caption{Comparisons of 2D pixel histograms for models A and D for $z$-bin pairs 1-3, 1-5 and 3-5, accompanied by the same 1D histograms from the top row of Figure \ref{fig:real_vs_gan_pix_n_peak}. The inner and outer contours represent the 68\% and 95\% one-sided quantiles respectively.}
    \label{fig:real_vs_gan_contour_pixel}
\end{figure}

\begin{figure}
    \centering
    \includegraphics[width=1\textwidth]{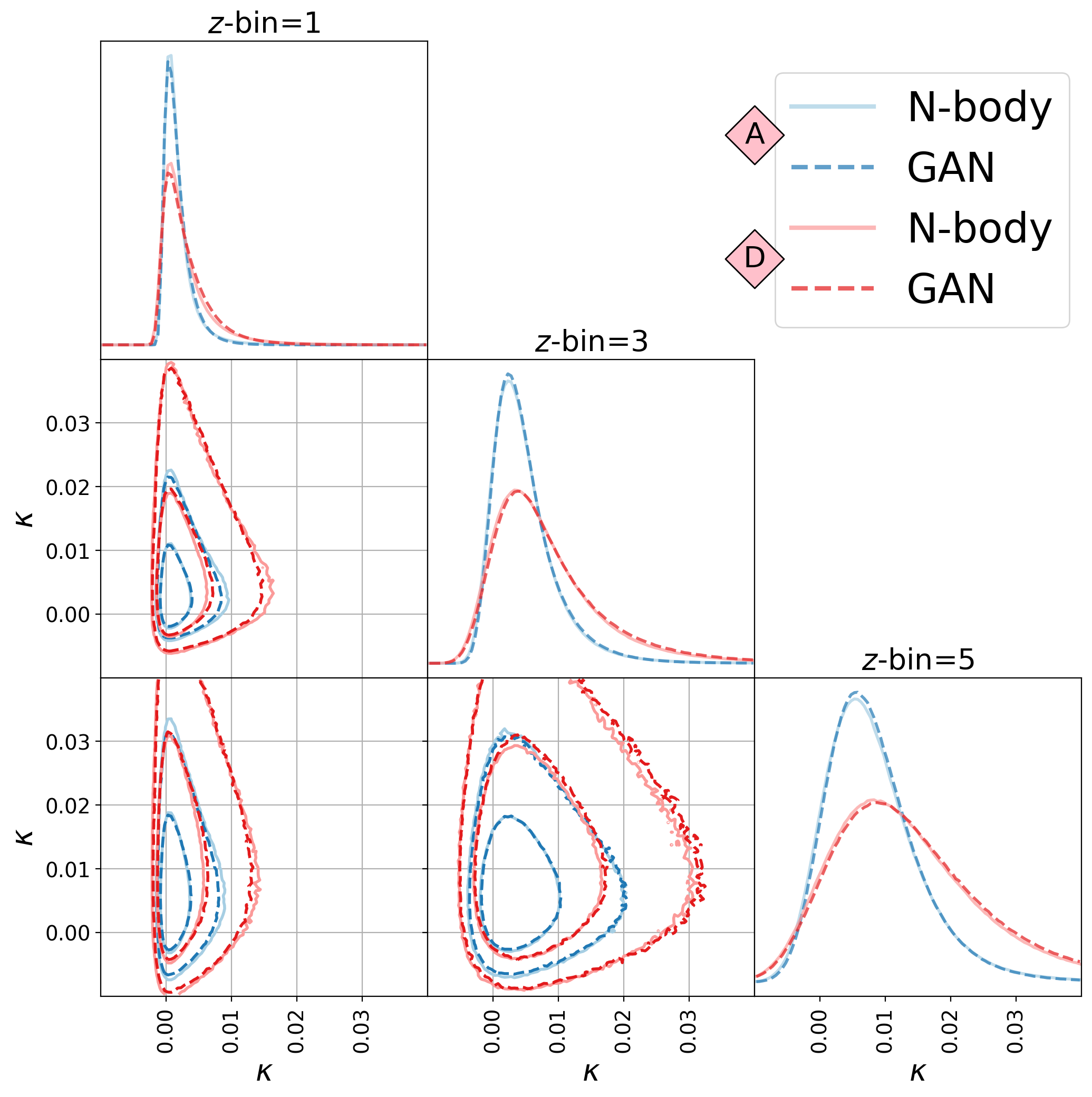}
    \caption{Same as Figure \ref{fig:real_vs_gan_contour_pixel}, but with comparisons of 2D peak histograms instead.}
    \label{fig:real_vs_gan_contour_peak}
\end{figure}

Similarly, the agreement between the peak histograms follows the same trend as that between the pixel histograms; the Hellinger distances of the peak values distribution $H_{\textrm{peaks}}$ for model A are $0.0445, 0.0351, 0.0401, 0.0418, 0.0388$ for tomographic bins 1 to 5 respectively, and the overall Hellinger distances of the peak distribution $H_{\textrm{peaks}}$ averaged over all tomographic bins are $0.0401, 0.0417, 0.0445, 0.0434$ for models A, B, C, D respectively.
Again, the GAN model follows the peak statistics better at higher redshifts than at lower ones. Although the disagreement in terms of the peak distribution $H_{\textrm{peaks}}$ is slightly larger than that of the pixel distribution $H_{\textrm{pixels}}$, that is expected since the presence of peaks is much rarer.

In order to show the joint distribution of pixel values in the maps from different redshifts, we show the 2D version of the pixel histograms and peak histograms in Figures~\ref{fig:real_vs_gan_contour_pixel} and \ref{fig:real_vs_gan_contour_peak} respectively, for the two test models A and D.
Again, we see an overall good agreement between the generated and original pixel/peak distributions. 
There are small noticeable differences in joint pixel distribution of $z$-bins 1 and 3.

\begin{figure}
    \centering
    \includegraphics[width=1\textwidth]{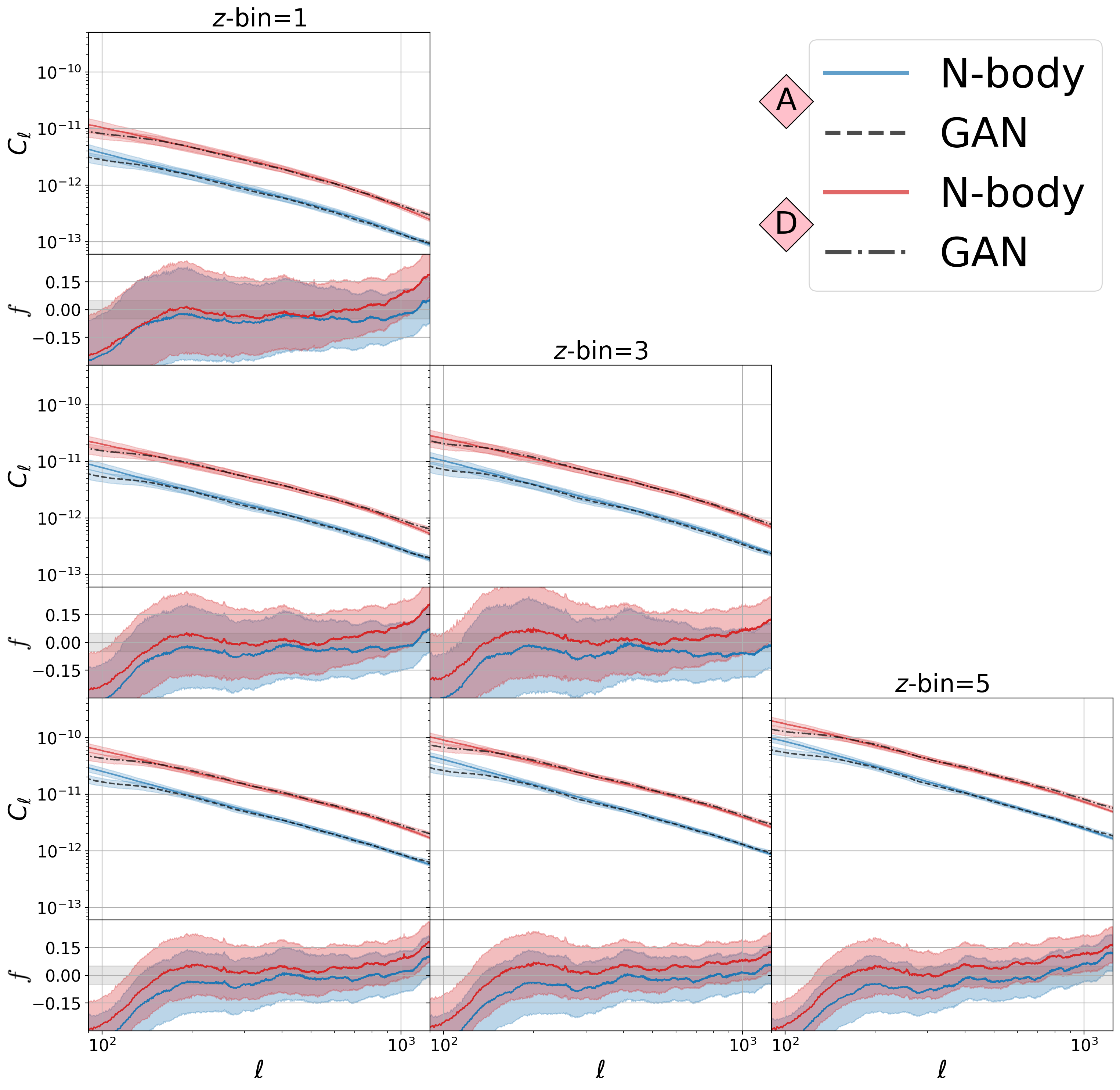}
    \caption{Comparisons of the angular power spectra $C_\ell$ for models A and D for $z$-bins$=1,3,5$, and the cross spectra for $z$-bin pairs 1-3, 1-5 and 3-5. The structure of this figure follows that of Figure \ref{fig:real_vs_gan_contour_pixel}, with the addition of a bottom panel for each spectra plot showing the fractional difference $f$ between the mean N-body and GAN power spectra. Both $C_\ell$ and $\ell$ axes are in logarithmic scales. As usual, each coloured band corresponds to one standard deviation of the respective data.}
    \label{fig:real_vs_gan_power}
\end{figure}

Figure \ref{fig:real_vs_gan_power} shows the angular power spectra $C_\ell$ of the real data and that of the generated ones are in good agreement; the two curves overlap with each other very well, with slight deviations in the low $\ell$ region. Specifically, the overall fractional differences between the N-body and GAN angular power spectra averaged over all $z$-bins and $z$-bin pairs throughout the range $\ell \in [100,1000]$ are: $f_{C_\ell}=4.28\%, 2.84\%, 5.85\%, 4.41\%$ for models A, B, C, D respectively. However, the GAN model underestimates $C_\ell$ at low $\ell$ in particular, exceeding the $10\%$ difference for $\ell \lesssim 150$.
One possible reason for such deviation at low $\ell$ would be cosmic variance; the total size of the KiDS-1000 footprint is still quite small relative to the full sky (accounting for $< 10\%$ of the sky), and thus the structures in one region of the sky may differ substantially from that in another region. Therefore, the power spectra of different training samples vary greatly, and so the model learnt to reflect this cosmic variance as well. Still, the means should not differ so significantly simply due to cosmic variance.
Another possible explanation for the deviation considers the nature of our graph-based spherical GAN model; by construction, the graph convolution layers from \deepsphere\ only allow pixels to interact within their local neighbourhood \cite{defferrard2020deepsphere}. Since the KiDS-1000 footprint consists of 3 disconnected regions, these regions fail to exchange information efficiently when passing through the GAN model.
In deep learning context, this implies that the effective receptive field \cite{Luo2017receptive} of the neural network is too small for the GAN to learn the large-scale statistics of the map. Therefore, the model has difficulty with reproducing the angular power spectrum at large scales. 
We expect simply connected footprints to perform better for our model, or novel approaches using transformer architectures \cite{Dosovitskiy2020transformer,Dwivedi2020transformer} might be able to solve these issues in future work.


\begin{figure}
    \centering
    \includegraphics[width=1\textwidth]{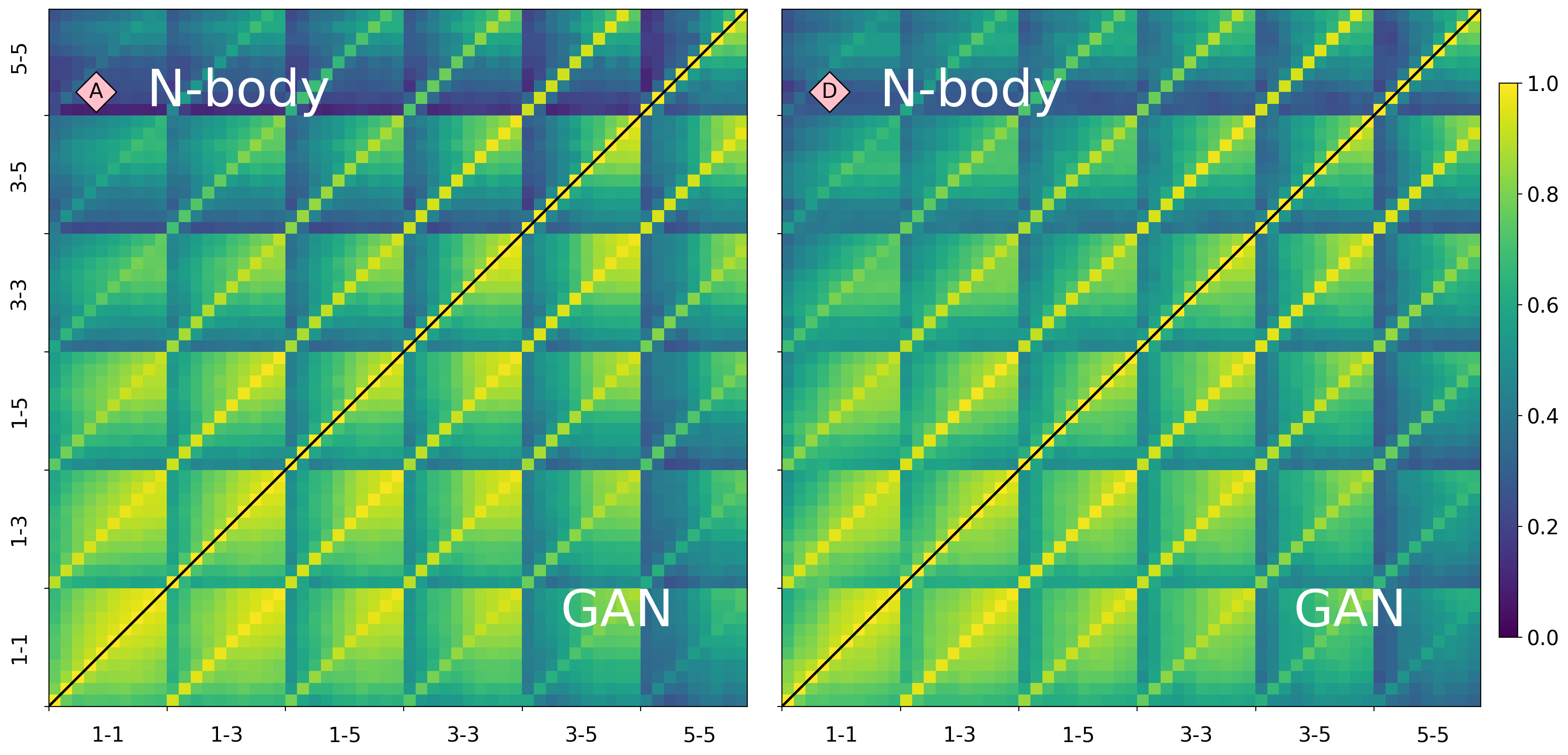}
    \caption{Pearson's correlation matrices $R_{\ell\ell'}$ for models A and D. This includes the auto-correlation for $z$-bins 1-1, 3-3, 5-5 \& $z$-bin pairs 1-3, 1-5 and 3-5, and the cross-correlation between them.}
    \label{fig:real_vs_gan_cov}
\end{figure}

The Pearson’s correlation matrices $R_{\ell\ell'}$ of the power spectra are shown in Figure \ref{fig:real_vs_gan_cov}. These correlations are created from 10 logarithmic bins in the $\ell \in [100, 1000]$ for each auto- and cross-correlation. The upper left and lower right triangular parts of each matrix show the original N-body correlations and the GAN correlations respectively. The Frobenius norms of these matrices were computed and the agreement was quantified using Equation \ref{eq:frobenius}. The fractional differences for models A, B, C, D are: $f_R = 0.134, 0.187, 0.070, 0.067$ respectively. The agreement is overall good, with models A and B being slightly worse. The correlation matrices for models B and C are shown in Appendix \ref{app:additional_plots}.
For intuitive interpretation of this difference, we note that for a single unit multivariate Gaussian, a $1\sigma$ difference in diagonal elements would lead to $f_R=1$. Therefore, this implies that the agreement between the N-body and GAN correlation matrices is on the level of $\sim15\%$.


\begin{figure}
    \centering
    \includegraphics[width=.9\textwidth]{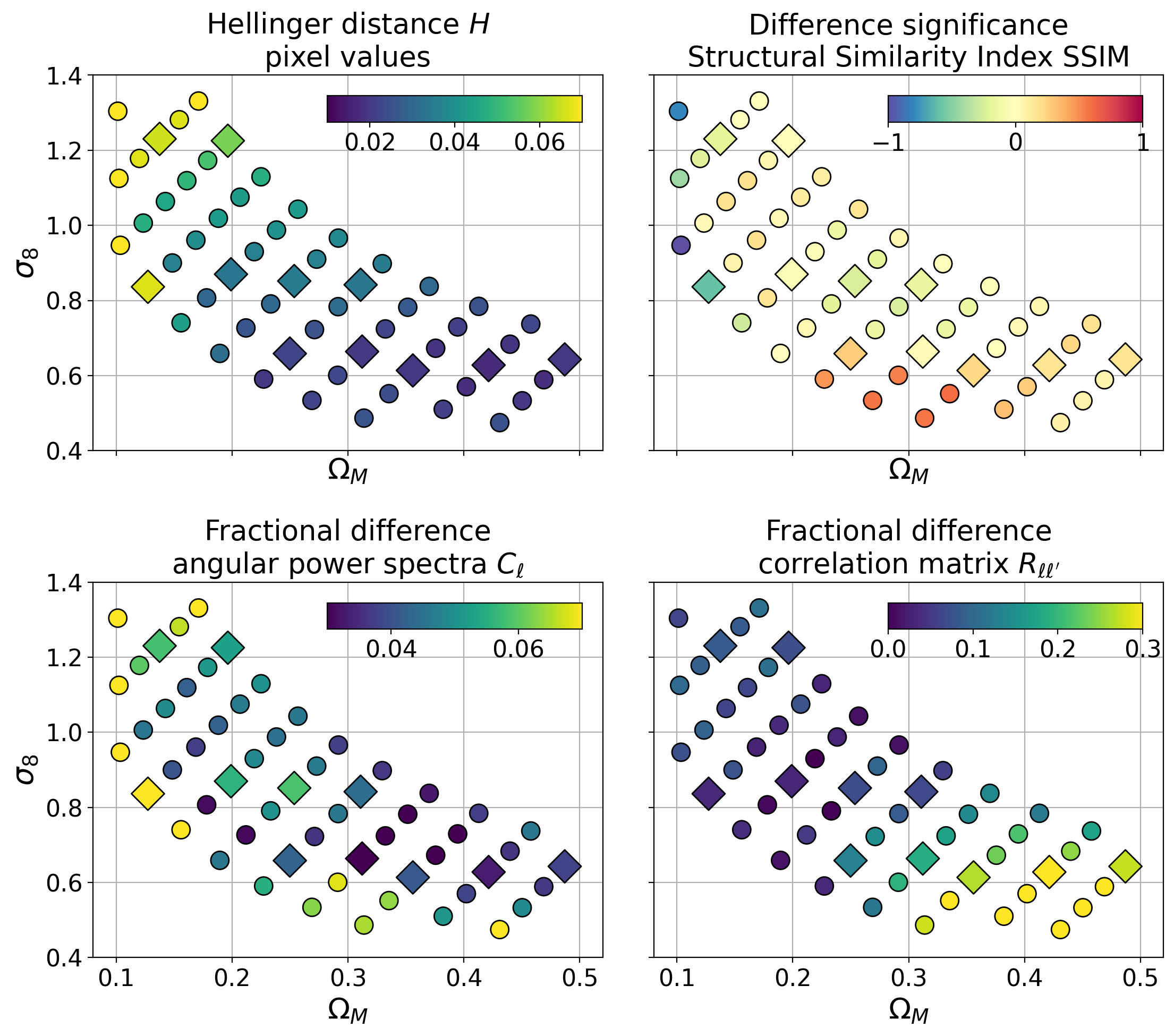}
    \caption{Differences between the various summary statistics of the original N-body and GAN-generated samples for different cosmologies. The Hellinger distance (Eq.~\ref{eq:hellinger_simple}) and difference significance of the SSIM Index (Eq.~\ref{eq:sig_diff}) are averaged over all 5 $z$-bins.
    Fractional differences in $C_\ell$ (Eq.~\ref{eq:psfracdiff}) and fractional differences in correlation matrices $R_{\ell\ell'}$ (Eq.~\ref{eq:frobenius}) are shown with an absolute value. Fractional differences in $C_\ell$ are averaged over all 5 auto-correlations plus the 10 cross-correlations.
    The difference metrics are constructed so that, for a Gaussian, a $1\sigma$ difference in mean or variance would lead to the difference of $\sim$1.
    Therefore, this Figure indicates that the agreement is of order <$10\%$ error for the centre of the grid, and slightly degrades towards the edges.}
    \label{fig:real_vs_gan_all}
\end{figure}

As the SSIM function from Equation \ref{eq:ssim} takes in two images, we would choose 500 random pairs of the GAN-generated samples of a particular cosmology and compute the SSIM scores. As each sample consists of 5 tomographic maps, we obtained 5 SSIM scores for each pair combination. They were averaged only at the very end of the computation. Therefore, by averaging the SSIM scores of these 500 pairs, the mean and the standard deviation of the SSIM for GAN-generated samples are obtained. These values would then be compared to the original N-body SSIM, which are computed from 500 randomly selected pairs of real samples as well, with the agreement of SSIM quantified by the difference significance formula described in Equation \ref{eq:sig_diff}. Such procedure is repeated for every cosmology, both from training and test set. In particular, the SSIM significance differences for models A, B, C, D are: $s_{\textrm{SSIM}} = 0.271, 0.034, -0.237, -0.162$. Overall, this indicates that the difference between the SSIM Index is much smaller than 1$\sigma$ in general, and thus the real data and generated data are in very good agreement in terms of SSIM scores. This result is consistent with visual inspection, where no mode collapse was found.


Finally, the comparisons of the aforementioned summary statistics between the original N-body and GAN samples are computed as a function of cosmology governed by the two conditioning parameters $(\Omega_M, \sigma_8)$ for both the training sets and test sets, and the plots are shown in Figure \ref{fig:real_vs_gan_all}.
They include the average Hellinger distance of the pixel values distribution, the average significance of the difference in the SSIM Index, the average absolute fractional difference in the angular power spectra $C_\ell$, and the absolute fractional difference in the Frobenius norm of the Pearson's correlation matrix $R_{\ell\ell'}$. The former two statistics are averaged over the 5 redshift bins, whereas the $C_\ell$ statistics averaged over all 5 auto- and 10 cross-spectra. 
Overall, there are no signs of overtraining; the model's performance does not seem to depend on whether the cosmology belongs to the training set or test set. Coupled with the fact that the statistical agreement of one particular cosmology is similar to that of its neighbouring cosmologies, this indicates that the emulator has learnt the latent interpolation of the maps efficiently.

Moreover, we can see in Figure \ref{fig:real_vs_gan_all} that the agreement between the N-body-simulated and GAN-generated maps in summary statistics is best at the centre of the grid for both the training and test set, and gets slightly worse as we move to the boundary of the grid. For example, the Hellinger distance is $H_{\textrm{pixels}} \lesssim 0.04$ for the majority of the cosmologies, with the very few exceptions at the leftmost edge of the grid, where the value of $H_{\textrm{pixels}}$ exceeds 0.06. Similarly, the difference significance of the SSIM Index for most cosmologies are very close to 0, with the outliers at the grid edges.
The fractional differences in the angular power spectra $C_\ell$ show similar results as well; the agreement of $C_\ell$ is $\sim 5\%$, with the centre performing slightly better while the edges performing slightly worse.
As for the correlation matrices $R_{\ell\ell'}$ of the angular power spectra, the quality deterioration is the most prominent at the bottom right edge of the grid, for high~$\Omega_M$ and low~$\sigma_8$.

\subsection{Cosmological constraints}
\label{sec:constraints}

The final test performed is on the level of mock cosmological constraints obtained from the original and emulated mass maps.
The maps are contaminated with a random Gaussian noise on the level corresponding to the KiDS-1000 survey.
We create the constraints using three summary statistics: angular power spectra, peak counts, and mass map histograms.
We perform the analysis using two approaches: 
(i) parametric likelihood, where we represent the likelihood as a multivariate Gaussian, as it is commonly done in this type of studies, and
(ii) likelihood via density estimation, using the \textsc{PyDelfi} package.
The details of the calculation of the constraints can be found in Appendix~\ref{app:constraints}.

\begin{figure}
    \centering
    \includegraphics[width=0.49\textwidth]{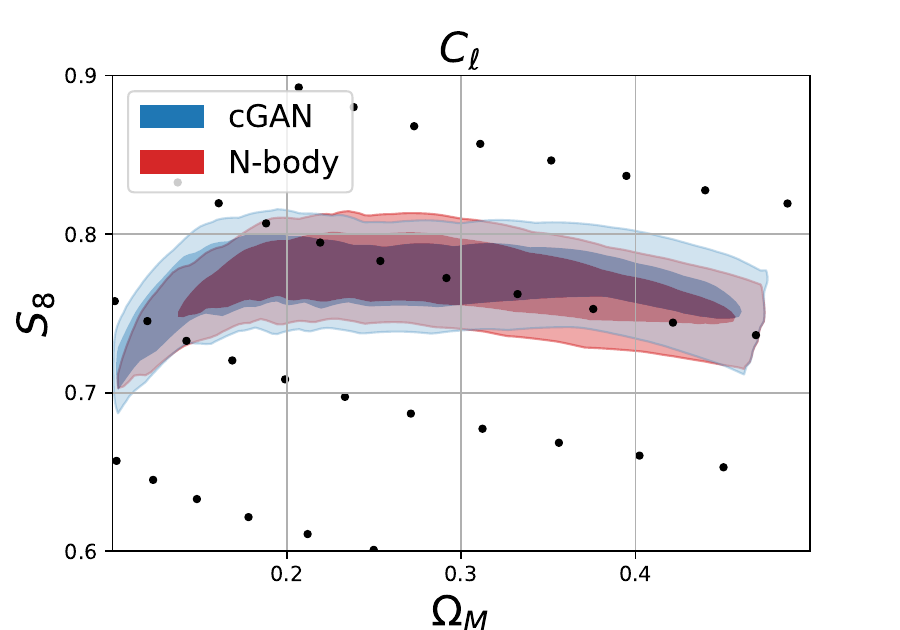}
    \includegraphics[width=0.49\textwidth]{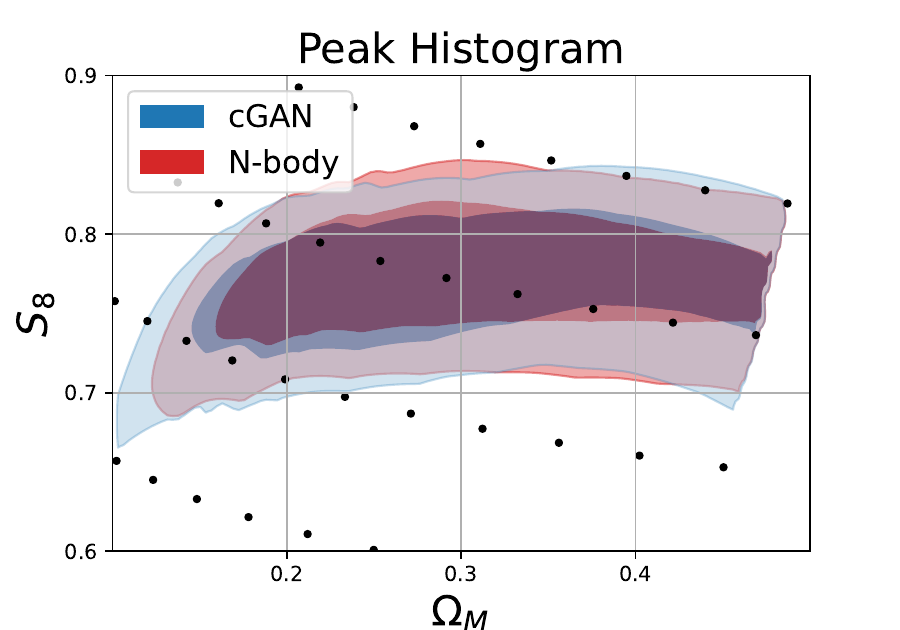}
    \includegraphics[width=0.49\textwidth]{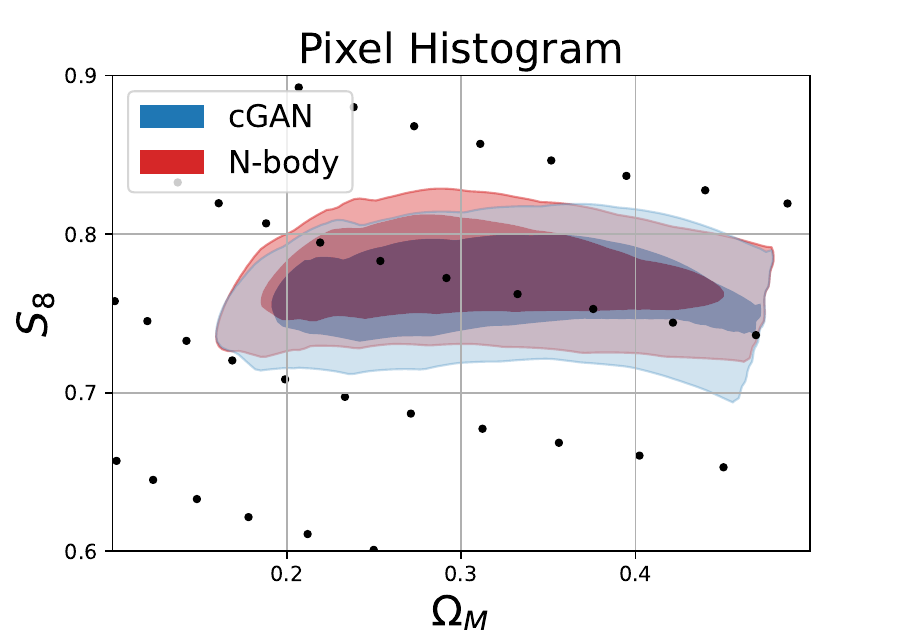}
    \includegraphics[width=0.49\textwidth]{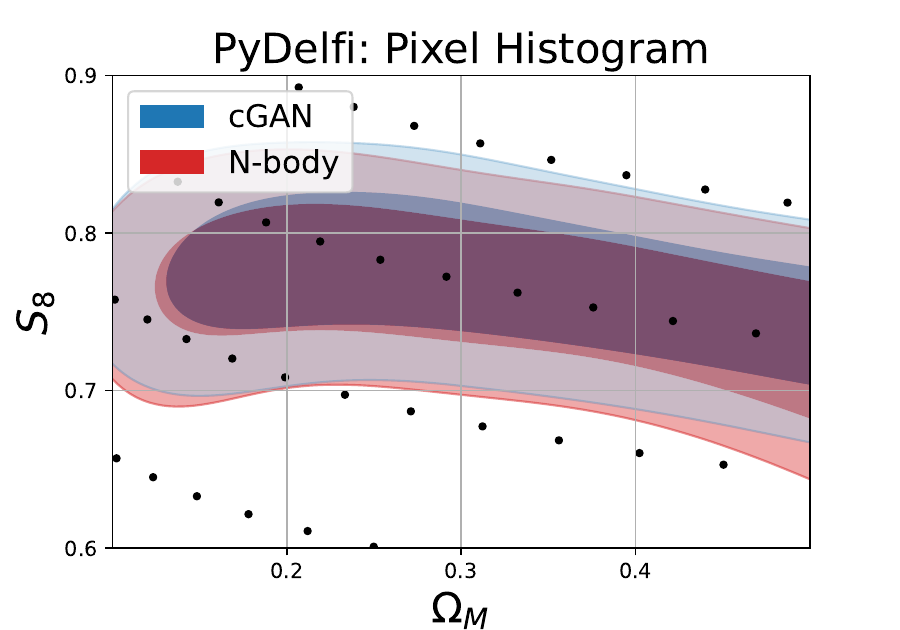}
    \caption{Parameter constraints on $\Omega_M$ and the degeneracy parameter $S_8 \equiv \sigma_8\sqrt{\Omega_M/0.3}$ using a mock observation and predictions from the N-body simulations or the GAN. The lower right panel shows the results of the \textsc{PyDelfi} histogram as summary statistics, while using the GAN and the original N-body simulation grid. The black dots indicate the simulated cosmologies. Note that the right bound of the constraints is an artefact of the interpolation scheme that was used in the likelihood analysis, as the interpolation of the data vector beyond the convex hull of the grid points is not possible. \label{fig:constraints}}
\end{figure}

The constraints of our analysis are shown in Figure \ref{fig:constraints} and presented in numerical form in Table \ref{tab:constraints}.
We perform a likelihood analysis as described in Appendix \ref{app:constraints} using the described summary statistics as data vectors. For the pixel and peak histograms, we use 10 linearly spaced bins between $-0.15$ and $0.15$ for all redshift bins individually, but only keep the bins that contained at least 25 counts over all cosmologies. For the power spectra, we use 9 logarithmic bins in the $\ell$-range 100 to 1000 for both auto- and cross-correlations. The constraints are generated using the mean summary of our fiducial simulations as mock observation. Otherwise, all data are solely from N-body simulations or the cGAN. The constraints generally agree well and the confidence intervals, provided in Table \ref{tab:constraints}, are less than one standard deviation apart. The statistical uncertainty of the parameters is therefore larger than the bias introduced by the cGAN. Additionally, one should note that the likelihood used in the analysis depends on the estimated summaries and their uncertainties. We therefore expect a small difference for finite sample sizes. The remaining small difference is caused by the fact that our cGAN does not perfectly reproduce the distribution of the simulations. Furthermore, including systematic effects would broaden the constraints even more, thus reducing the relative difference between the two approaches. Finally, we calculate the Jensen-Shannon divergence (JSD) of the posterior distributions obtained with the simulations and the cGAN. The JSD is a distance measure between probability distributions similar to the Wasserstein distance, but based on the Kullback–Leibler divergence. A distance of 0 indicates that the distributions are identical and the upper bound of the JSD is 1. All calculated values are also reported in Table~\ref{tab:constraints}. This small difference is comparable to the shifts in the posterior distribution of \cite{kids1000_shear} introduced by different modelling choices of the systematics. This shows that the cGAN produces maps that are accurate enough to constrain cosmological parameters for a KiDS-like survey. We note, however, that one should proceed with caution regarding using the first redshift bin, larger scales, and cosmology point on the edge of the grid. Moreover, more investigation would be required for other types of summary statistics than shown here.

To obtain the \textsc{PyDelfi} constraints, we used the pixel histogram as the summary statistic.
The \textsc{PyDelfi} constraints are broader than the ones obtained with the Gaussian likelihood function, which is not unexpected. A reason for this difference is the fact that \textsc{PyDelfi} does not make any assumptions about the distribution of the summary statistic, whereas the others explicitly model the likelihood as a Gaussian distribution. For example, using a Gaussian likelihood with a fixed covariance matrix can lead to underestimated constraints~\cite{fluri2018cosmological}, increasing the difference with the constraints.
This difference could be decreased if a cosmology-dependent covariance matrix was used \citep{Eifler2009covariance}.
As the goal of this paper is to assess the quality of the match between the cGAN and the N-body simulations, we leave the investigation of the choices of summary statistic and likelihood-building methods to future work.


\section{Conclusion}
\label{sec:conclusions}

We implemented a tomographic spherical mass map emulator for the KiDS-1000 survey using a Wasserstein generative adversarial network, based on spherical convolutional neural networks in \deepsphere. We conditioned the model on the cosmological parameters $\Omega_M$ and $\sigma_8$. 
We train the model on the tomographic mass maps produced by N-body simulations.
The map emulator can then sample new tomographic mass maps for a given cosmology; the sampled maps are statistically very close to the N-body simulated ones.
We have demonstrated the capability of the model to perform latent interpolation efficiently and give realistic draws for any cosmology, including those that were not used for training.

\begin{table}
    \centering
    \begin{tabular}{llrrrr}
         \hline
         Summary & Constraint & $\Omega_M$ & $\sigma_8$ & $S_8$ & JSD \\
         \hline
         \hline
         \multirow{2}{*}{Power Spectra} & N-body & $0.28 \pm 0.09$ & $0.84 \pm 0.15$ & $0.77 \pm 0.02$ & \multirow{2}{*}{0.044} \\
                                           & GAN & $0.25 \pm 0.10$ & $0.90 \pm 0.18$ & $0.77 \pm 0.02$ & \\
         \hline
         \multirow{2}{*}{Peak Histogram} & N-body & $0.29 \pm 0.09$ & $0.81 \pm 0.13$ & $0.77 \pm 0.03$ & \multirow{2}{*}{0.01}\\
                                           & GAN & $0.28 \pm 0.09$ & $0.82 \pm 0.14$ & $0.77 \pm 0.03$ & \\
         \hline
         \multirow{2}{*}{Pixel Histogram} & N-body & $0.30 \pm 0.07$ & $0.79 \pm 0.10$ & $0.77 \pm 0.02$ & \multirow{2}{*}{0.03} \\
                                           & GAN & $0.31 \pm 0.08$ & $0.77 \pm 0.10$ & $0.76 \pm 0.02$ & \\
         \hline
    \end{tabular}
    \caption{Mean and standard deviation of the parameter constraints obtained using either the N-body simulations or the cGAN. The Jensen-Shannon divergence (JSD) is a distance measure between the two distributions and was calculated using the 2D $\Omega_M$-$\sigma_8$ constraints. \label{tab:constraints}}
\end{table}

We evaluate the performance of the emulator using multiple quantitative metrics.
In particular, the agreements of the non-tomographic summary statistics (pixel histograms, angular power spectra, peak histograms, and correlation matrices) are very good, with the former two typically on the $\lesssim 5\%$ level, and the latter two on the $\sim 10\%$ level. Likewise, the tomographic statistics (cross power spectra, 2D pixel/peak histograms, and cross-correlation matrices) generally agree well, indicating that the model has effectively learned the tomographic map information.
The comparison of the Structural Similarity (SSIM) Index also shows a good agreement in this metric; the overall difference significance in SSIM is much smaller than $1\sigma$ for most cosmologies, with the very few exceptions of the left edge (low $\Omega_M$, high $\sigma_8$) of the grid. This indicates the absence of model overfitting.
However, the quality of the map is particularly worse at low redshift bins and at high negative $\kappa$ values, which signifies that the emulator is particularly weak in following the statistics of underdense regions of the map. We inform the readers to consider these potential weaknesses when using this emulator with summary statistic that are very sensitive to these observables.

Furthermore, our GAN model is able to capture the variability in the dataset as a function of $\Omega_m$ and $\sigma_8$: the agreements in the summary statistics between the original and generated samples are good throughout all cosmologies, from both the training and test sets. In general, the performance of the model is the best at the centre of the grid and only deteriorates when reaching the edges. Specifically, the agreement in correlation matrices $R_{\ell\ell'}$ worsens for the high $\Omega_M$ $-$ low $\sigma_8$ region, whereas that in the other summary statistics worsens for the low $\Omega_M$ $-$ high $\sigma_8$ edge. Such deterioration is understandable, as the training set contains less information near the edges of the grid for the GAN to learn the latent interpolation effectively. Therefore, more investigation is required to improve the performance of the generative model in those regions. These inaccuracies of the GAN model is most likely not due to the lack of training data, as we originally only had 250 mock surveys instead of 1000 for each cosmology, and yet the increase in training data did not yield significant improvement. One reason could be due to the architecture as we did not perform any rigorous hyperparameter search for computational reasons. But the more probable reason would be the intrinsic design of the spherical GAN that makes it particularly weak at dealing with disconnected regions due to the limited effective receptive field of the neural network. However, as generative models are growing in popularity in deep learning, we anticipate that all these aforementioned problems can be solved in the near future. For example, using cycleGANs \cite{Zhu2017cycleGAN} may improve the statistical agreement, and novel approaches using transformer architectures \cite{Dosovitskiy2020transformer,Dwivedi2020transformer} might be able to solve these receptive field issues in future work. For now, we simply advise the readers to avoid generating mass maps from those regions when using the model.

Our model generates new tomographic mass maps in a fraction of a second at any point in the $\Omega_m-\sigma_8$ space.
This can enable fast likelihood-free inference based on MCMC sampling, active learning, or other schemes that explore the parameter space in an adaptive way. Additionally, the generator is fully differentiable, making it for example possible to use methods like information maximising neural networks (IMNN)~\cite{Charnock2018} without relying on finite difference methods to estimate the derivatives. 
This can give a particular advantage when the parameter space is expanded to include systematics, such as intrinsic alignments, shear calibration errors, baryons, and others.

The emulator is publicly available and requires only \texttt{TensorFlow} and \texttt{healpy} packages.
It generates maps corresponding to the current state-of-the-art KiDS-1000 lensing dataset. Ease of access and installation of the emulator, as well as its speed, can lower the entry barrier for those interested in rapidly testing new ideas in the field of cosmological inference.

To the best of our knowledge, this is the first application of spherical GAN to emulate mass maps with a realistic footprint. Therefore, our results bring the idea of deep generative models to practical use in a full, end-to-end cosmological analysis closer than ever before. 
Future prospects include further improvements to the precision of the emulator, such that it can break through the $99\%$ agreement barrier and make the real and generated samples indistinguishable within the limits given by cosmic variance.
With the interest in machine learning and generative models continuing to reach an all-time high, we expect that to be achievable in the near future, as machine learning techniques as well as hardware will most certainly keep improving.
Another interesting direction is to expand the emulator with more fields, such as galaxy clustering, intrinsic alignment, and baryons.

\acknowledgments

We would like to thank Alexandre Refregier for the useful advice and helpful discussions. 
We also thank Nathana\"el Perraudin,  Aurelien Lucchi, and Sandro Marcon for building the foundations for this project.
We thank Tilman Tr\"oster and Catherine Heymans for useful comments.
As mentioned in Section \ref{sec:implementation}, in addition to \texttt{TensorFlow}\footnote{\url{https://www.tensorflow.org}} and \texttt{Keras}\footnote{\url{https://keras.io}}, we made use of the functionalities provided by \texttt{numpy}\footnote{\url{https://numpy.org}}, \texttt{scipy}\footnote{\url{https://scipy.org}}, \texttt{healpy}\footnote{\url{https://healpy.readthedocs.io/en/latest/}} and \texttt{matplotlib}\footnote{\url{https://matplotlib.org}}.

Based on observations made with ESO Telescopes at the La Silla Paranal Observatory under programme IDs 177.A-3016, 177.A-3017, 177.A-3018 and 179.A-2004, and on data products produced by the KiDS consortium. The KiDS production team acknowledges support from: Deutsche Forschungsgemeinschaft, ERC, NOVA and NWO-M grants; Target; the University of Padova, and the University Federico II (Naples).


\bibliography{library}
\bibliographystyle{ieeetr_short}

\appendix

\section{Qualitative comparison metrics}
\label{app:metrics_description}
This section describes the various summary statistics used in this work as the metrics for gauging the performance of the model, as mentioned in Section \ref{sec:metric}.

\subsection{Angular power spectrum} \label{app:power_spec}
The angular two-point correlation function fully encapsulates all Gaussian information of convergence maps \cite{zurcher2021}, making it the most important summary statistics. Its Fourier counterpart, the angular power spectrum, characterises the size of the fluctuations as a function of angular scale, i.e. it shows the frequency of periodic structures of a given size found in a map, with large-scale structures corresponding to small multipole moments $\ell$, and small-scale structures to large $\ell$. Let us consider some field $F(\hat{\theta})$ defined over the full sky. One can decompose $F(\hat{\theta})$ in spherical harmonics as
\begin{equation}
    F(\hat{\theta}) = \sum_{\ell=0}^{\infty}\sum_{m=-\ell}^{\ell}a_{\ell m}Y_{\ell m}(\hat{\theta}), \quad \quad a_{\ell m} \equiv \int d^2\Omega_{\hat{\theta}} Y^{*}_{\ell m}(\hat{\theta})F(\hat{\theta}),
\end{equation}
where $\hat{\theta}$ is the unit direction vector and $Y_{\ell m}(\hat{\theta})$ is the spherical harmonic function \cite{zurcher2021}. One can define the angular power spectrum $C_\ell$ using the relation
\begin{equation}
    \langle a_{\ell m} a^*_{\ell'm'} \rangle = \sum^l_{m=-\ell} \frac{|a_{\ell m}|^2}{2l+1}\delta_{\ell\ell'} \delta_{mm'} = C_\ell \delta_{\ell\ell'}\delta_{mm'},
\end{equation}
with the last equality assuming each $a_{\ell m}$ is a standard normal deviate \cite{hinshaw2003first}. Therefore, for a convergence field $\kappa$ with its divergence-free component neglected\footnote{In the general case both the curl-free $\kappa_E$ and the divergence-free $\kappa_B$ components are both considered, the angular power spectrum would end up with pure E-modes, pure B-modes, and the mixing term of EB-modes. But since E-modes are by far the strongest, the other terms are usually neglected.}, the angular power spectrum of $\kappa$ can be computed \cite{zurcher2021}.

\subsection{Mass map histogram and peak counts} \label{app:peak_hist}

As the pixel value of the convergence map corresponds to the mass integrated in that particular line of sight, this makes the mass map histograms a very simple yet useful statistics for comparing the maps \cite{original_cGAN} and constraining cosmological models (see Section \ref{app:constraints} for more details). Additionally, peak statistics has been shown to help capture non-Gaussian information in weak lensing surveys \cite{Berg__2010}, as highly non-linear structures such as clusters of galaxies (and hence massive dark matter haloes) engrave themselves onto convergence maps in the form of `peaks', i.e. local maxima \cite{zurcher2021}. The peak statistics counts the number of peaks in distinct value ranges. For the convergence maps, a pixel whose value is greater than that of its eight neighbouring pixels is defined as a peak. The peak statistics simply counts the number of peaks in each pixel intensity bin.

To quantify the agreement between the pixel/peak distribution of N-body-simulated and GAN-generated maps aside from histograms, \citetalias{original_cGAN} use the Wasserstein-1 distance $W_1(u , v)$. Also known as earth mover's distance, it informally describes minimum cost of transporting mass in converting the data distribution $u$ to the data distribution $v$ achieved with the optimal transport plan $\gamma$. Mathematically, it is defined as
\begin{equation}
\label{eq:original_wasserstein_inf}
    W_1(u, v)=\inf_{\gamma \in \Pi(u, v)} \int_{\mathbb{R}\times\mathbb{R}} \abs{x-y} d\gamma(x,y),
\end{equation}
where $\Pi(u, v)$ is the set of all joint distributions on $\mathbb{R}\times\mathbb{R}$ with marginals $u$ and $v$ on the first and second factors respectively, i.e. the set contains all possible transport plans $\gamma$. If $U$ and $V$ are the respective cumulative distribution functions of $u$ and $v$, then the distance can also be written as
\begin{equation}
\label{eq:original_wasserstein_scipy}
    W_1(u, v)=\int_{-\infty}^{+\infty}\abs{U-V}.
\end{equation}
See \cite{wassersteindualityproof} for proof of the equivalence of these two definitions. The $W_1$ computation by \texttt{scipy} follows these definitions.

However, since the Wasserstein distance is scale-dependent, the pixel values of the map must be normalised before calculating the distance. This is achieved by subtracting the mean $\mu$ from the map $m$ and then the map is divided by the standard deviation $\sigma$. We use $\mu$ and $\sigma$ of all original N-body maps for a given cosmology for normalising both N-body and GAN samples, i.e.
\begin{equation}
\label{eq:wasserstein}
    W_1^s = W_1 \left(  \frac{m_{\textrm{GAN}} - \mu_\textrm{N-body}}{\sigma_\textrm{N-body}}, \frac{m_{\textrm{N-body}}-\mu_\textrm{N-body}}{\sigma_\textrm{N-body}} \right),
\end{equation} 
where $\mu_\textrm{N-body}$ and $\sigma_\textrm{N-body}$ are the mean and standard deviation of all pixels in the N-body maps for a given cosmology respectively, and $W_1$ is the unscaled Wasserstein-1 distance (see Equation \ref{eq:original_wasserstein_inf} for definition).
Now, the $W_1$ distance can easily be interpreted; for a Gaussian distribution with $\mu=0$, $\sigma=1$, a $1\sigma$ shift in the mean implies $W_1=1$, whereas scaling its variance by a factor of 2 implies $W_1 \approx 0.8$ \cite{original_cGAN}. This way, we remove the dependence of the $W_1$ distance on the map dynamic range scale and enable quantitative comparison.

\subsection{Pearson correlation coefficients of $C_\ell$} \label{app:correlation_matrix}

The Pearson correlation coefficient matrix $R_{\ell\ell'}$ of the angular power spectrum $C_\ell$ is the normalised definition of the covariance matrix $\textrm{cov}(\ell,\ell')$, which is defined as a reduced cross product of $C_\ell$ at different $\ell$ for the same realisation averaged over different realisations \cite{Klypin2018correlation_matrices}. Mathematically,
\begin{equation}
    R_{\ell\ell'} \equiv \frac{\textrm{cov}(\ell,\ell')}{\sqrt{\textrm{cov}(\ell,\ell)\textrm{cov}(\ell',\ell')}},
\end{equation}
where
\begin{equation}
    \textrm{cov}(\ell,\ell') \equiv \langle C_\ell C_{\ell'} \rangle - \langle C_\ell \rangle  \langle C_{\ell'} \rangle.
\end{equation}
By definition, $R_{\ell\ell}=1$. Such quantity measures the degree of non-linearity and mode coupling of waves with different moments, thus playing an important role in estimates of the accuracy of measured power spectrum \cite{Klypin2018correlation_matrices}. This makes the correlation matrix an interesting metric for gauging the performance of the GAN.

\subsection{Unique version of SSIM} \label{app:ssim}

\begin{figure}
    \centering
    \includegraphics[width=.6\textwidth]{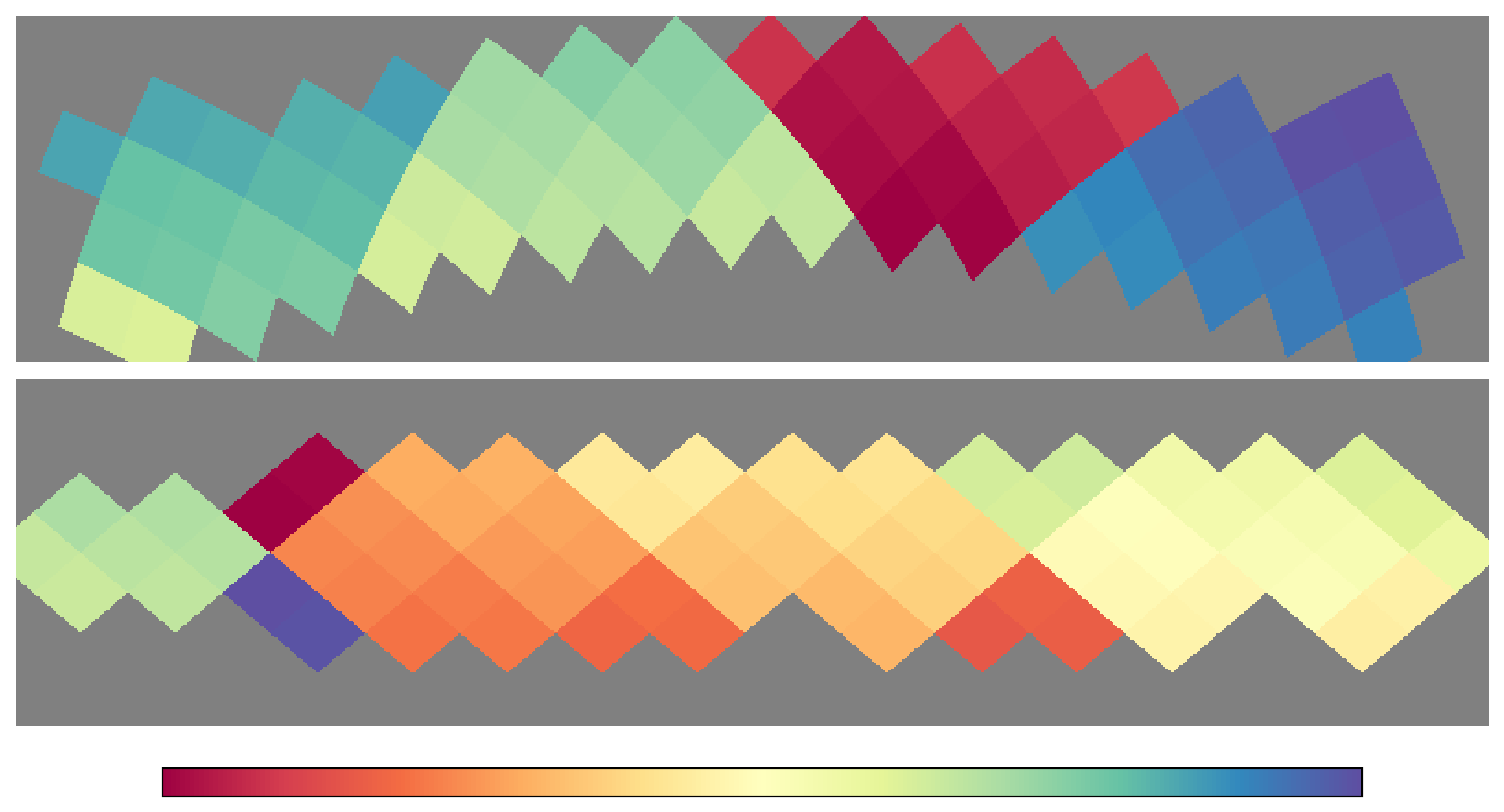}    \caption{Superpixels in the map. The size of the superpixel is $32^2=1024$, and the total number of pixels $N_{\textrm{pix}}$ of the map is 149504. The difference in colour between the superpixels corresponds to the separation in the 1D array. This is especially significant since the KiDS-1000 footprint consists of 3 disconnected regions.}
    \label{fig:superpixel}
\end{figure}

In addition to the summary statistics, there exists a popular method for image quality assessment in machine learning called the structural similarity (SSIM) index measure \cite{wang2004ssim}. Originally proposed for quantifying perceptual similarity, it has been adopted into a standard metric in neural networks \cite{zhao2017loss}. However, since our samples are now \texttt{HEALPix} maps in NESTED ordering represented by 1D-arrays rather than conventional 2D images, we would need to slightly adjust the defintion of SSIM in order to use it as one of the quantitative comparison metrics for the cGAN. The SSIM formula takes in two images $x, y$ and returns a score in the range $[0, 1]$, where 0 indicates no structure similarity in the two images, and 1 indicates identical images. Mathematically, it is defined as
\begin{equation} \label{eq:ssim}
    \textrm{SSIM}(x,y)=\frac{(2 \mu_x \mu_y + c_1)(2\sigma_{xy}+c_2)}{(\mu_x^2 + \mu_y^2 + c_1)(\sigma_x^2+\sigma_y^2+c_2)},
\end{equation}
where $\mu_{x,y}$ is the mean of $x$ or $y$, $\sigma_{x,y}^2$ is the variance of $x$ or $y$, $\sigma_{xy}$ is the covariance of $x$ and $y$, $c_1=(0.01L)^2$ and $c_2=(0.03L)^2$ are two small constants, and $L$ is the dynamic range of the pixel values \cite{wang2003multiscale}. Here, we choose the value of $L$ for each tomographic bin such that it includes $99.7\%$ of the pixels, i.e. $L=0.0164, 0.0278, 0.0383, 0.0534, 0.0629$ for $z$-bins 1 to 5.
Therefore, in the context of GAN, the SSIM index is a good indication of mode collapse; we can use this method to measure the similarities in a set of generated maps, ideally the SSIM index of the generated data would be as low as that of the real data, implying the generated maps are as diverse as the real maps.
Typically, the SSIM index is calculated by sliding a window pixel-by-pixel across the whole image space \cite{wang2003multiscale}. However, with the nature of \texttt{HEALPix} NESTED ordering and the fact that the map has disconnected patches, pixels far apart geometrically may be neighbouring each other in the 1D array representing the map, and vice versa. To prevent this, we propose a different approach: the size of the window is chosen to be that of a superpixel such that it can evenly divide the number of pixels $N_{\textrm{pix}}$ of the sample (see Figure \ref{fig:superpixel}). Now, instead of the sliding window method, we simply compute the SSIM of every superpixel and take the average. This way, the pixels within the window would always be geometrically close to each other during the SSIM computation.


\section{Cosmological Constraints}
\label{app:constraints}

In addition to having accurate summary statistics, it is important that the generated maps can be used to infer cosmological parameters. Given a mock observation, we constrain the cosmological parameters $\Omega_M$ and $\sigma_8$ using either the simulated (ground truth) or the generated convergence maps as signal. However, as opposed to the other comparison metrics, we add a realistic amount of observational noise to the maps. We mimic realistic observational noise using a Gaussian random field with zero mean and variance \cite{obs_noise}
\begin{equation}
    \sigma^2 = \frac{\sigma_e^2}{An_g},
\end{equation}
where $\sigma_e = 0.3$ is a common ellipticity dispersion, $A$ is the pixel area and $n_g$ is the observed galaxy density. We set $n_g = 1.5$ galaxies/arcmin$^2$ for each redshift bin, leading to a total galaxy density of $n_g^\mathrm{tot} = 7.5$ galaxies/arcmin$^2$, which is similar to the observed galaxy density in the KiDS-1000 data set \cite{kids1000_shear}.

\subsection{Multivariate Gaussian Likelihood}
\label{app:gauss_cons}

To constrain the cosmological parameters, we perform a parameter inference similar to \cite{kids450_dl}. We use the Gaussian likelihood described in \cite{Jeffrey2018} that is based on \cite{Sellentin2015} and includes the uncertainty of the estimated means and covariance matrices. Given a covariance matrix $\hat{S}$ that is obtained from $N$ simulated data vectors and an estimated mean $\hat{\mu}$ from $M$ simulated data vectors, the likelihood of observing a particular data vector $d$ is given by
\begin{equation}
    P(d\vert\hat{\mu},M,\hat{S},N)\propto \vert\hat{S}\vert^{-\frac{1}{2}}\left( 1 + \frac{M}{(M+1)(N-1)}\mathbf{Q} \right)^{-\frac{N}{2}}, \label{eq:likelihood}
\end{equation}
where
\begin{equation}
    \mathbf{Q} = (d - \hat{\mu})^T\hat{S}(d-\hat{\mu}),
\end{equation}
and the normalisation constant is irrelevant for our purposes. We estimate the covariance matrix $\hat{S}$ using our summaries as data vectors
\begin{equation}
    \hat{S} = \frac{1}{N-1}\sum_{i = 1}^N (d_i - \hat{\mu})(d_i - \hat{\mu})^T.
\end{equation}
We make the common assumption that the covariance matrix does not depend on the cosmological parameters and estimate it using the $N = 2500$ realisations of our fiducial cosmology or the same number of realisations generated with our cGAN respectively. As in \cite{kids450_dl}, we estimate the means $\hat{\mu}_i$ for each grid point using our $M = 1000$ simulated or generated realisations. Afterwards, we linearly interpolate the estimated means across our priors and evaluate the likelihood of equation \eqref{eq:likelihood} on a regular grid and calculate the posterior distribution using Bayes theorem,
\begin{equation}
    P(\Omega_M, \sigma_8 \vert d_\mathrm{obs}) = \frac{P(d\vert\hat{\mu},M,\hat{S},N)p(\Omega_M,\sigma_8)}{p(d_\mathrm{obs})}\propto P(d\vert\hat{\mu},M,\hat{S},N) ,
\end{equation}
where $p(\Omega_M,\sigma_8)$ is our flat prior ($\Omega_M \in [0.1, 0.4]$ and $\sigma_8 \in [0.5, 1.4]$) and the evidence $p(d_\mathrm{obs})$ can be absorbed into the normalisation. 

\subsection{Likelihood density estimation with \textsc{PyDelfi}}
\label{app:ABC_cons}

One of the main advantages of the cGAN is that it can be sampled for arbitrary parameter choices $(\Omega_M,\sigma_8)$, making it possible to use other inference techniques than the one presented in \cite{kids450_dl}. This is especially useful for likelihood free inference (LFI) methods, where an explicit parametric expression for the likelihood, like equation \eqref{eq:likelihood}, is not necessary. Instead, these methods use density estimation to build the likelihood function. There already exist applicable and advanced LFI methods like \textsc{PyDelfi} \cite{Alsing2019pydelfi} or \textsc{GPABC} (Gaussian Process Approximate Bayesian Computation \cite{gpabc}). 



In this work, we used \textsc{PyDelfi}. 
It works by learning the distribution of the data $p(d\vert\theta)$ given a vector of cosmological parameters $\theta$ using masked auto-regressive flows and mixture density networks \cite{Alsing2019pydelfi}. We train these networks on the same data that was used for the likelihood analysis of the pixel histograms and perform an inference. However, we reduced the dimensionality of the data from 50 to six by performing an singular value decomposition and only keeping the six most important coefficients. This was done because \textsc{PyDelfi} can be unstable for high-dimensional distributions. 
We performed the \textsc{PyDelfi} analysis on both the cGAN emulator and the original simulation grid points.
The results are shown in the lower right panel in Figure~\ref{fig:constraints}.
It can be seen that the discrepancy of the two \textsc{PyDelfi} runs are similar to the discrepancies of the likelihood analyses shown in section \ref{sec:results}. 
The overall size of the constraints is larger than for the multivariate Gaussian; this is most likely due to the fact that the covariance matrix was calculated only at the fiducial point, rather than varying across the parameter space. 
It is known that results obtained using a cosmology-dependent covariance matrix can differ from fixed covariance \cite{Eifler2009covariance, fluri2018cosmological}.  


\section{Network architecture}
\label{app:architecture}

The full architecture of the spherical cGAN is shown in Table \ref{table:archi}. There, the batch size $b$ is defined as the number of samples passed through each training step, `conv' denotes a strided convolutional layer from DeepSphere, `deconv' denotes the transposed version, and `Chebyshev' denotes the graph convolution layer with Chebyshev \cite{deepsphere} polynomials of degree $K=5$, with a $k=20$-nearest neighbours graph \cite{defferrard2020deepsphere}. Additionally, $g_7$ is the layer responsible for forcing the pixels in the padded region to take the value of zero, $d_0$ is the (non-trainable) rescaling layer where the pixels of each of the 5 tomographic maps are rescaled such that their standard deviations are the same, and $d_{6,7}$ are the layers that flatten the tensor into a vector and then concatenate the conditioning parameters to it. As the training samples are 1D arrays with length (number of pixels $N_\textrm{pixels}$) 149504 and consist of 5 tomographic maps, the generator output and discriminator input were fixed to this dimension. The latent (input) dimension of the generator was 584 plus the 2 conditioning parameters ($\Omega_M, \sigma_8)$. One should note that this input dimension is much smaller than the final amount of pixels, also limiting the number of truly statistical independent pixel which could have an impact on complicated summary statistics with a dimension greater than 584. However, we did not observe such an effect on the summaries considered in this work. There were a total of $\sim 5$ million trainable parameters for each network in the model. Unlike normal 2D images, the pixel values of the patches were not bounded, so the generator output has no activation (i.e. linear).

The important hyperparameters not mentioned in Table \ref{table:archi} are as follow: the batch size $b$ was fixed to be 128, the initializer for all generator and discriminator layers was set to the default \verb|`glorot_uniform'| \cite{glorot2010glorot}, the optimiser for both $G$ and $D$ to be \texttt{Nadam} \cite{dozat2016nadam} with an initial learning rate of $l=\num{2e-4}$ and an exponential decay rate for the 1st moment estimates $\beta_1=0.5$ and 2nd moment $\beta_2=0.9$, and the penalty coefficient $\lambda$ was set to 10. The learning rate was later decreased $l=\num{2e-5}$ at epoch 350 to further stabilise training. The discriminator was updated 5 times more than the generator. Note that the batches were composed of samples from different cosmologies from the training set. The model was trained for 600 epochs, which corresponded to approximately 160 hours (around a week) on the Piz Daint cluster, running on a single GPU node.

\begin{table}
\centering
\begin{tabular}{ l c c c }
 \ChangeRT{1pt}
 Layer & Operation & Activation & Dimension \\ 
 \ChangeRT{1pt}
 \vspace*{-5pt} \\
 \textit{Generator} \\
 \cline{1-3}
 \vspace*{-5pt} \\
 \textbf{z} & input & linear & $b \times (584+2)$ \\
 
 $g_0$ & linear & ReLU & $b \times 9344$ \\
 $g_1$ & reshape & linear & $b \times 584 \times 16$ \\
 $g_2$ & Chebyshev & ReLU & $b \times 584 \times 64$ \\ 
 $g_3$ & deconv & ReLU & $b \times 2336 \times 64$ \\ 
 $g_4$ & deconv & ReLU & $b \times 9344 \times 32$ \\ 
 $g_5$ & deconv & ReLU & $b \times 37376 \times 16$ \\ 
 $g_6$ & deconv & linear & $b \times 149504 \times 5$ \\
 $g_7$ & lambda & linear & $b \times 149504 \times 5$ \\
 \vspace*{-5pt} \\
 \textit{Discriminator} \\
 \cline{1-3}
 \vspace*{-5pt} \\
 \textbf{X} & input & linear & $b \times 149504 \times 5$ \\
 $d_0$ & rescaling & linear & $b \times 149504 \times 5$ \\
 $d_1$ & conv & ReLU & $b \times 37376 \times 16$ \\ 
 $d_2$ & conv & ReLU & $b \times 9344 \times 32$ \\
 $d_3$ & conv & ReLU & $b \times 2336 \times 64$ \\
 $d_4$ & conv & ReLU & $b \times 584 \times 64$ \\ 
 $d_5$ & Chebyshev & ReLU & $b \times 584 \times 16$ \\
 $d_6$ & flatten & linear & $b \times 9344$ \\  
 $d_7$ & concatenate & linear & $b \times (9344+2)$ \\  
 $d_8$ & linear & ReLU & $b \times 584$ \\ 
 $d_{10}$ & linear & linear & $b \times 1$ \\ 

\end{tabular}

\caption{Conditional GAN architecture used in this work.} 
\label{table:archi}
\end{table}

\section{Complementary results}
\label{app:additional_plots}

The following Figures are additional result plots not shown in Section \ref{sec:results} for the sake of clarity. They should be self-explanatory after reading the main text.

\begin{figure}
    \centering
    \includegraphics[width=1\textwidth]{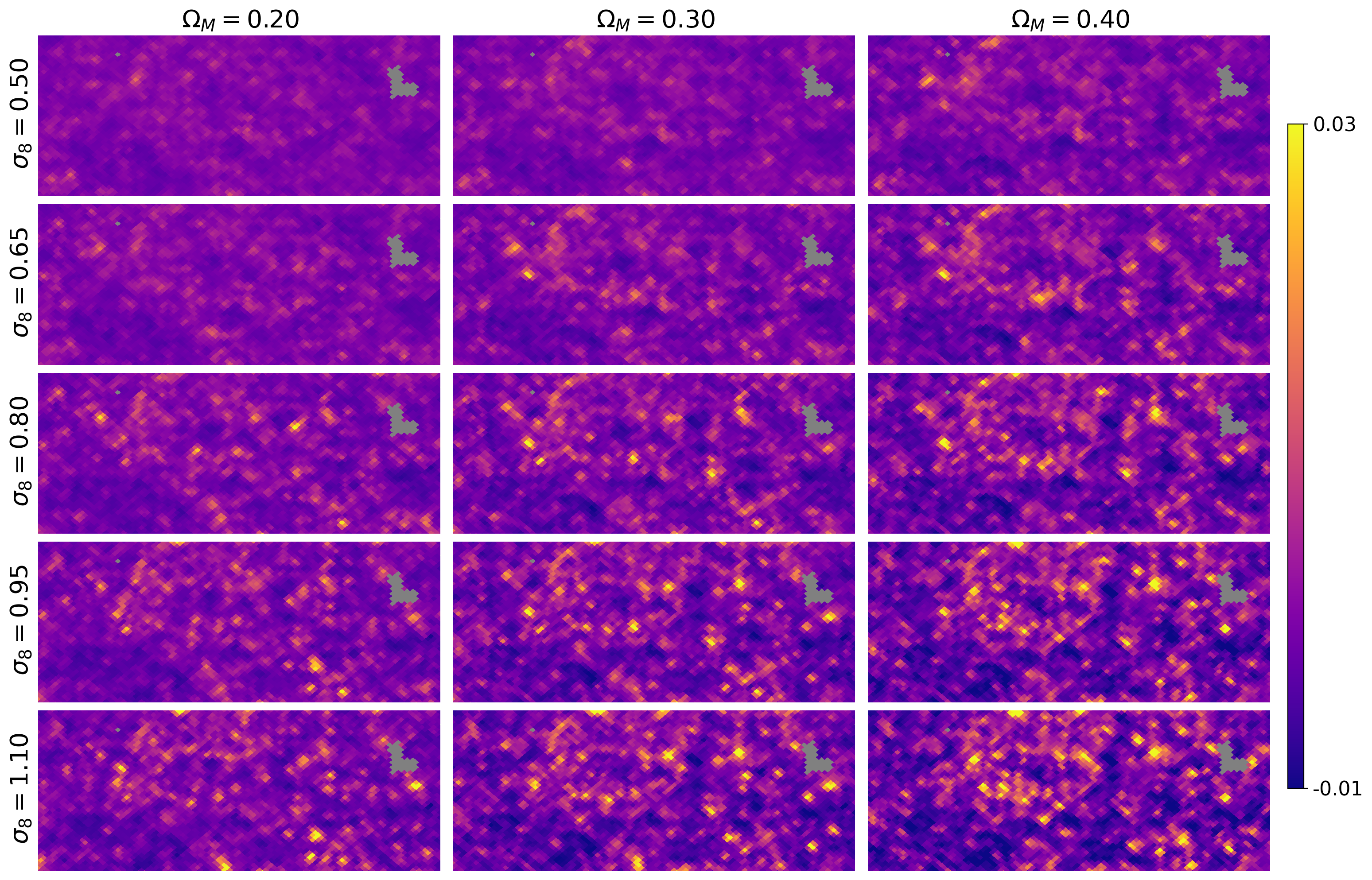}
    \caption{Generated samples from the same random seed but with varying cosmological parameters $\Omega_M$ and $\sigma_8$, showing only $z$-bin$=3$. This illustrate the smooth transitions of the generated maps from one cosmology to another, suggesting that the generator was successfully conditioned to these two parameters. The difference in generated convergence maps with different cosmologies agrees with theory as well; an increasing $\Omega_M$ leads to a higher mass density in the map in general, while an increasing $\sigma_8$ to a higher variance in pixel intensity.}
    \label{fig:latent_interpolation}
\end{figure}

\begin{figure}
    \centering
    \includegraphics[width=1\textwidth]{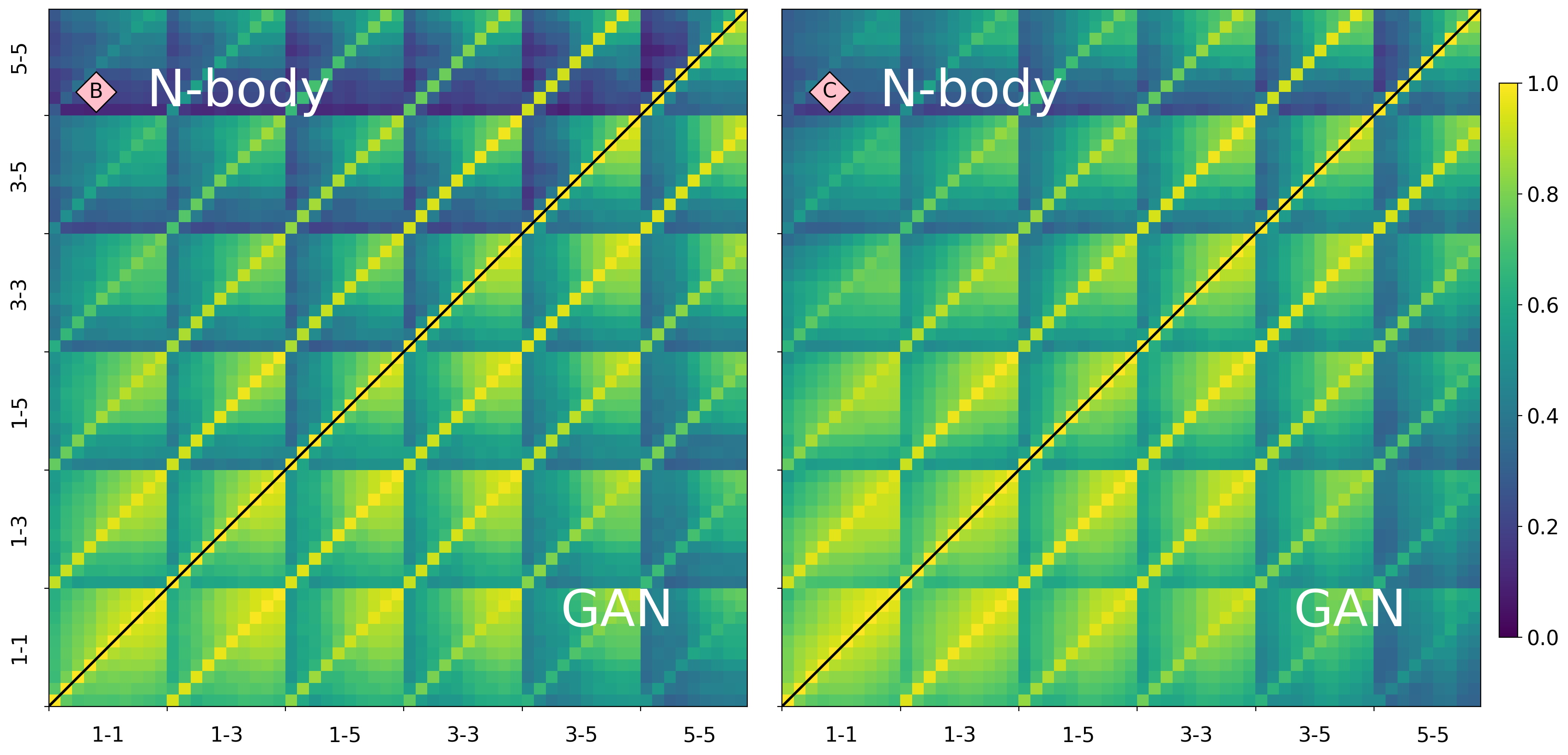}
    \caption{Pearson's correlation matrices $R_{\ell\ell'}$ for models B and C.}
    \label{fig:BC_corr}
\end{figure}

\begin{figure}
    \centering
    \includegraphics[width=1\textwidth]{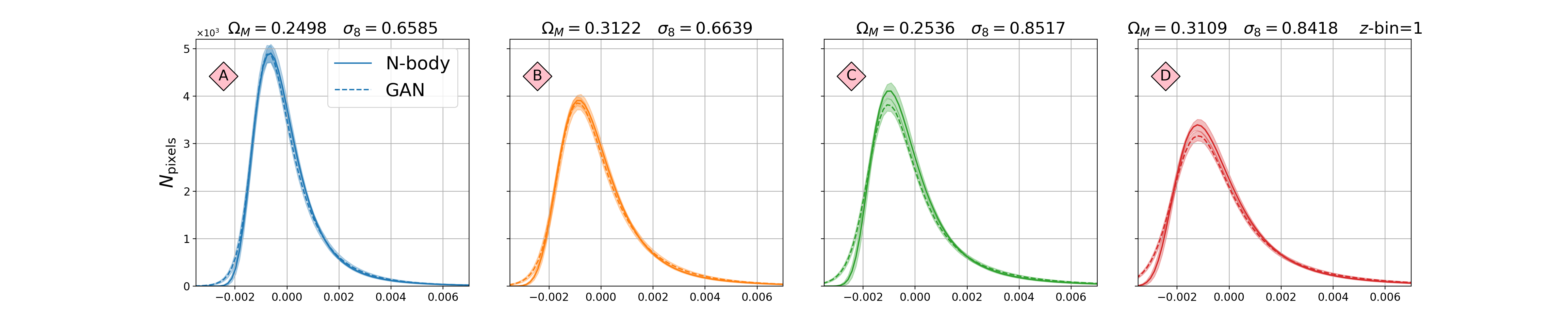}
    \includegraphics[width=1\textwidth]{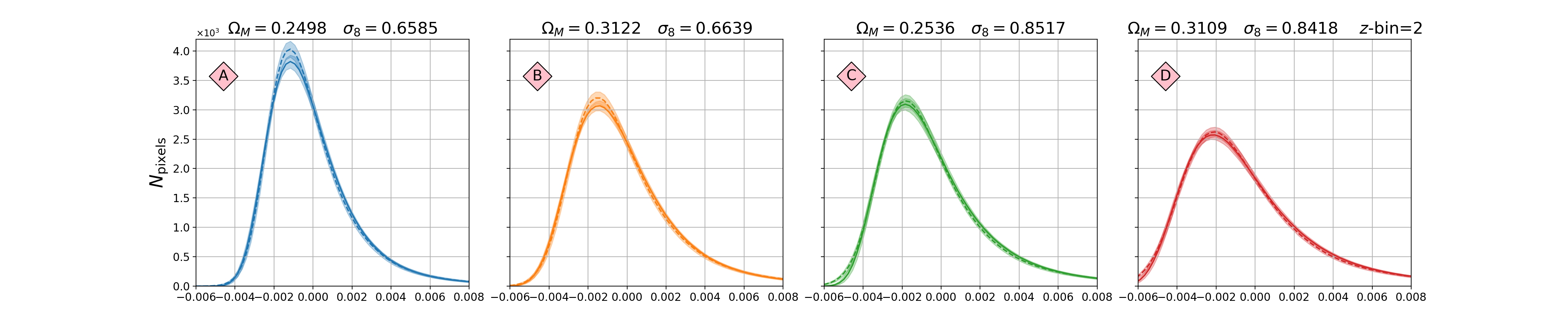}
    \includegraphics[width=1\textwidth]{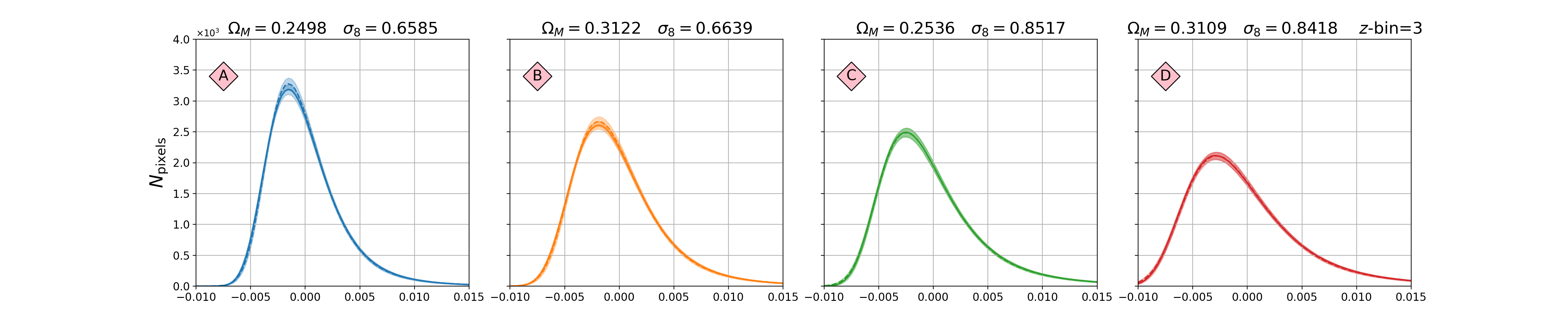}
    \includegraphics[width=1\textwidth]{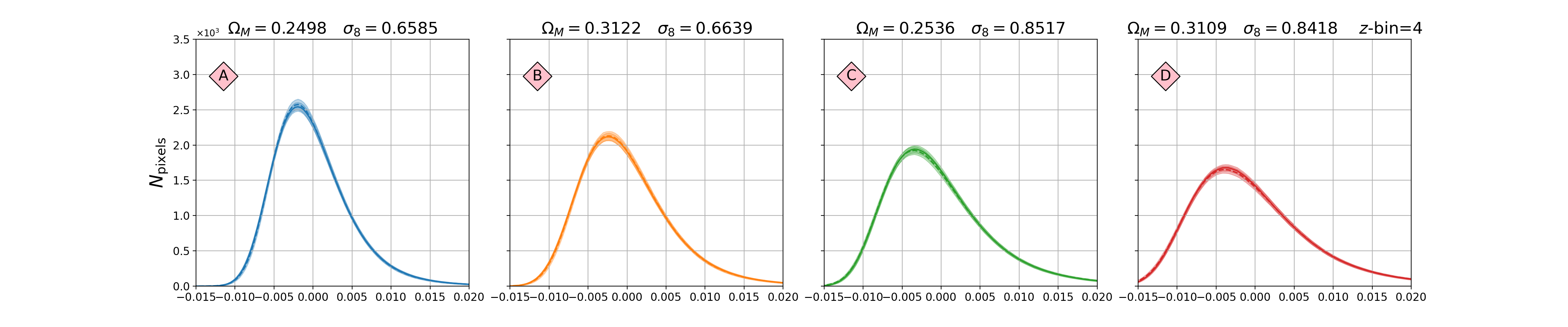}
    \includegraphics[width=1\textwidth]{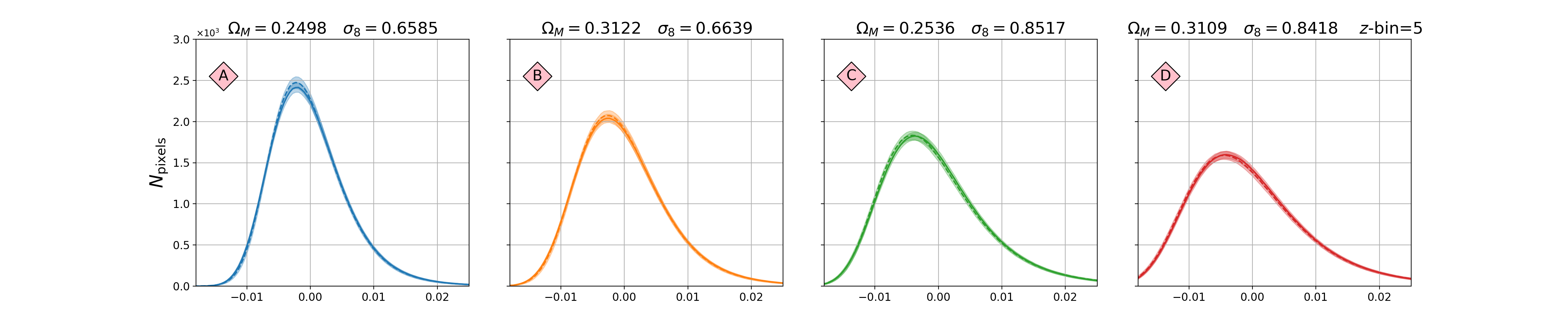}
    \caption{All tomographic mass map histograms for models A, B, C and D.}
    \label{fig:all_mass_hist}
\end{figure}

\begin{figure}
    \centering
    \includegraphics[width=1\textwidth]{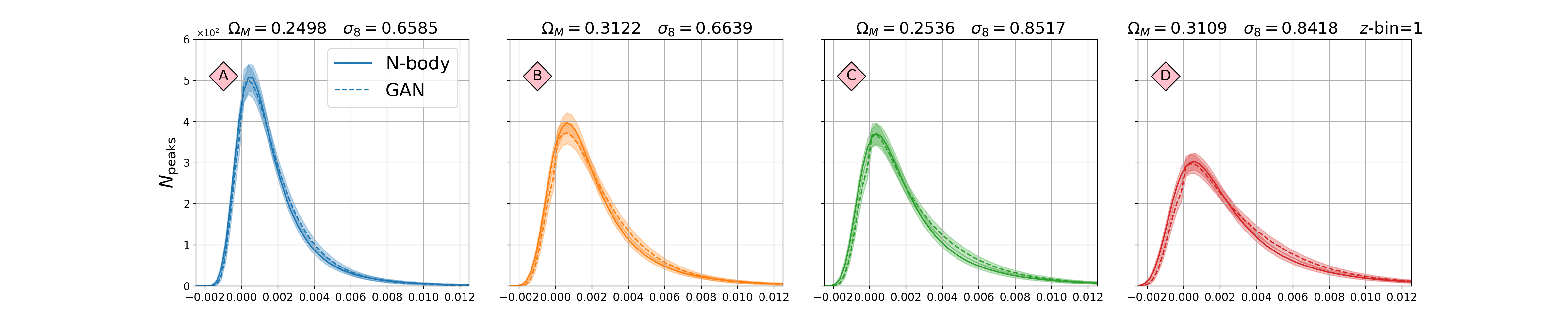}
    \includegraphics[width=1\textwidth]{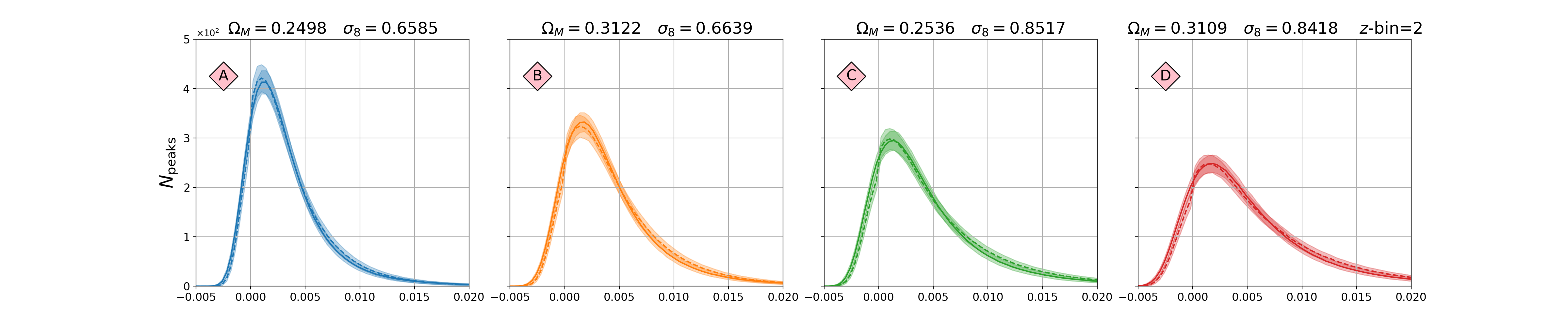}
    \includegraphics[width=1\textwidth]{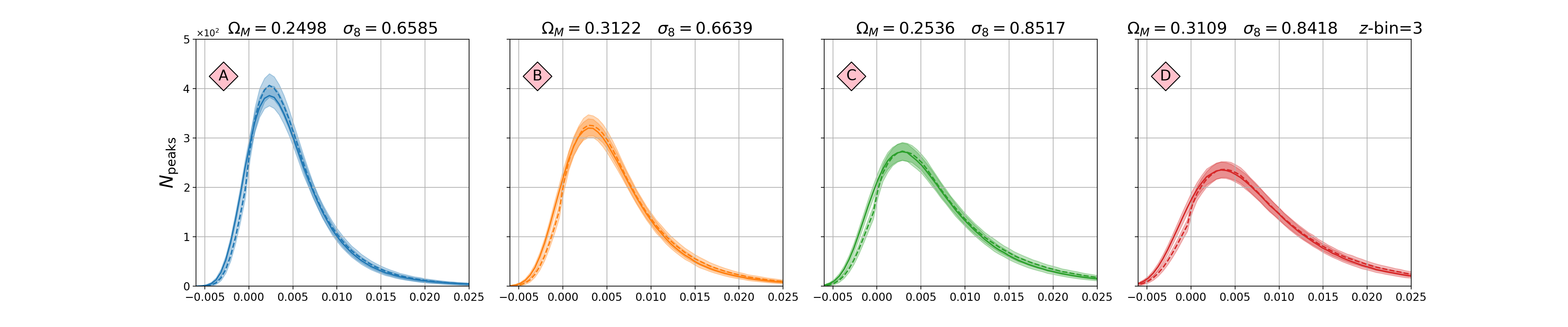}
    \includegraphics[width=1\textwidth]{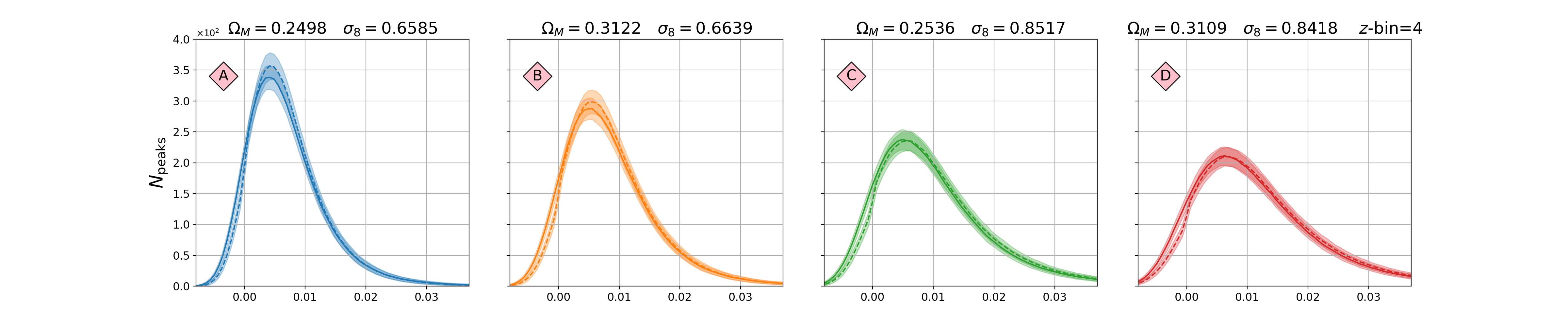}
    \includegraphics[width=1\textwidth]{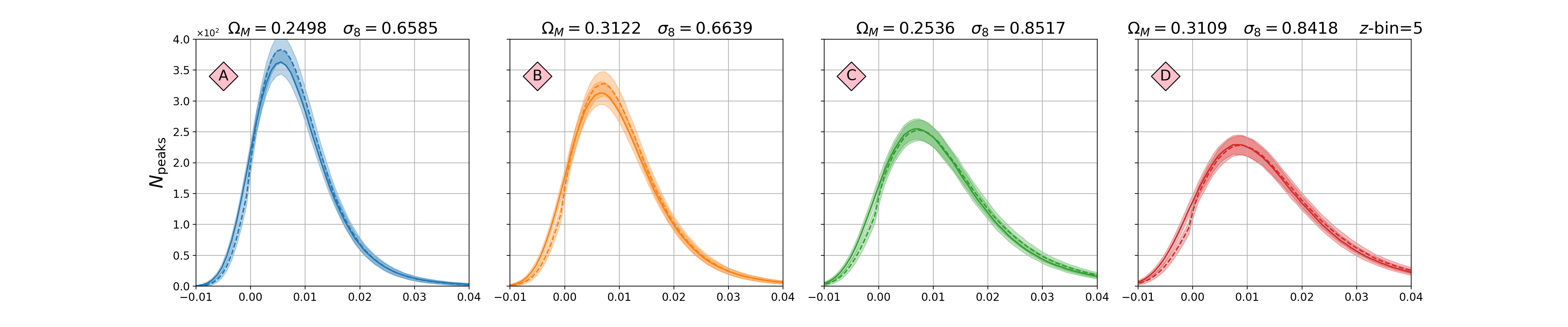}
    \caption{All tomographic peak histograms for models A, B, C and D.}
    \label{fig:all_peak_hist}
\end{figure}

\end{document}